\documentclass[prd,aps,twocolumn,a4paper,floatfix,nofootinbib,showpacs]{revtex4-1}

\usepackage{graphicx,psfrag}
\usepackage{mathrsfs}
\usepackage{amsmath,amsfonts,amssymb}
\usepackage{multirow}
\usepackage{comment}
\usepackage[colorlinks=true]{hyperref}
\hypersetup{
  linkcolor=blue,
  citecolor=blue,
  urlcolor=blue
}
\usepackage{bbm}
\usepackage{placeins}
\usepackage[normalem]{ulem}
\usepackage{tensor} %

\def\Lell{L}
\def\eps{\epsilon}

\newcommand{\eg}{e.g.\ }
\newcommand{\ie}{i.e.\ }

\usepackage{color}
\definecolor{cyan}{rgb}{0,0.9,0.9}
\definecolor{orange}{rgb}{0.9,0.5,0}
\definecolor{purple}{rgb}{0.8,0.4,0.8}
\definecolor{gray}{rgb}{0.8242,0.8242,0.8242}
\definecolor{grey}{rgb}{0.5,0.5,0.5}
\definecolor{pink}{rgb}{1.0, 0.0, 0.5}
\newcommand{\mytext}[1]{}

\renewcommand{\mytext}{}

\begin{document}

\title{Hyperbolic Relaxation Method for Elliptic Equations} 

\author{Hannes R. \surname{R\"uter}$^{1}$, David \surname{Hilditch}$^{1,2}$, Marcus
\surname{Bugner}$^{1}$, and Bernd \surname{Br\"ugmann}$^{1}$}

\affiliation{
${}^1$Theoretical Physics Institute, University of Jena, 07743 Jena, Germany.\\
${}^2$CENTRA, University of Lisbon, 1049 Lisboa, Portugal.
}

\date{\today}

\begin{abstract} 
We show how the basic idea of parabolic Jacobi relaxation can be
modified to obtain a new class of hyperbolic relaxation schemes that
are suitable for the solution of elliptic equations. Some of the
analytic and numerical properties of hyperbolic relaxation are
examined. We describe its implementation as a first order system in a
pseudospectral evolution code, demonstrating that certain elliptic
equations can be solved within a framework for hyperbolic evolution
systems. Applications include various initial data problems in
numerical general relativity. In particular we generate initial data
for the evolution of a massless scalar field, a single neutron star,
and binary neutron star systems.
\end{abstract}

\pacs{
  02.60.Cb,    %
  02.70.Hm,    %
  04.25.D-     %
}

\maketitle

\tableofcontents

\section{Introduction}\label{Section:Introduction}

The solution of elliptic partial differential equations (elliptic
PDEs) is an important problem in many areas of physics, including
fluid dynamics, quantum mechanics, general relativity, and many more
besides. Correspondingly large is the variety of analytic and
numerical methods dealing with the solution of elliptic PDEs. The
starting point for many methods is a discretization and (if required)
a linearization, which for typical problems arising in physics leads
to a sparse system of linear equations for a large but finite number
of degrees of freedom.

A key role in the solution of linear systems is played by iterative
methods, \eg \cite{Saa03,SaaVor00}. Among the basic iterative methods
are relaxation methods, in particular Gauss-Seidel and Jacobi
relaxation methods, and the family of Krylov subspace methods. Closely
connected are strategies to accelerate the convergence of such
methods, such as preconditioners and multigrid methods.

In this paper we study a modification of the classic Jacobi method,
which is motivated by its origin in physical relaxation problems. For
concreteness, consider as a minimal example the Laplace equation,
\begin{equation}
\Delta \phi = 0,
\label{laplace}
\end{equation}
for a function $\phi(x,y,z)$ on a regular subset of
$\mathbb{R}^3$ together with appropriate boundary conditions. The
Jacobi relaxation scheme can be obtained by introducing a time
parameter $t$ and considering instead of (\ref{laplace}) the parabolic
diffusion equation
\begin{equation} \partial_t \phi = \Delta \phi.
\label{diffusion}
\end{equation}
As time approaches infinity, any initial data for $\phi$ ``relaxes''
to a stationary state, where $\partial_t \phi = 0$ and hence
$\Delta\phi=0$ (\ref{laplace}) is satisfied as well. The Jacobi
iteration method is obtained by discretizing the diffusion equation
(\ref{diffusion}). In essence, we introduce an ``unphysical'' time
dependence, which is not part of the original problem, and obtain the
solution to the time-independent problem by means of a fixed point
iteration.

In this paper we investigate a similar strategy, which however relies
on a different type of evolution equation. Instead of replacing the
elliptic equation (\ref{laplace}) by the parabolic equation
(\ref{diffusion}), we consider a {\em hyperbolic} wave equation with
damping,
\begin{equation}
\partial_t^2 \phi + \partial_t \phi = \Delta \phi.
\label{wavediffusion}
\end{equation}
Combining time derivatives as in (\ref{wavediffusion}) adds
strong diffusion to the pure wave equation while maintaining the
hyperbolic character of the PDE. The idea is that deviations from the
stationary state satisfying $\Delta\phi=0$ are damped to zero or are
propagated away, and furthermore it can be advantageous to perform
hyperbolic as opposed to parabolic evolutions.

In the limit of vanishing damping we obtain the wave equation
\begin{equation}
  \partial_t^2 \phi = \Delta \phi.
  \label{wave}
\end{equation}
If a stationary state is reached, we again have
solved~(\ref{laplace}). Experimenting with (\ref{wavediffusion}) we
found that the damping is the main desirable feature, while
propagating waves off the grid is far less relevant for the reduction
of the residual.  Therefore, the proposal is to study
(\ref{wavediffusion}) for a variety of elliptic operators with strong
damping.

Hyperbolic equations containing diffusion or viscosity terms are a
well-known topic, \eg \cite{HsiLiu92,Nis97,Wir14}, which is relevant
to our discussion, see Sec.~\ref{Subsection:EvolutionSystem}. The
topic revolves around applying relaxation methods to hyperbolic
equations, in particular hyperbolic conservation laws with diffusion.
Although the type of equations that are considered are similar, our
perspective is different.  The elliptic equation is for us the
fundamental problem, and we add a hyperbolic, damped time-dependence
to obtain an iterative scheme for the solution of the elliptic
equation.  Since there does not seem to be an established name for
this idea, we refer to the method as {\em hyperbolic relaxation for
  elliptic equations} (HypRelax), as opposed to parabolic relaxation
that is at the heart of the Jacobi method.

With regard to previous literature on hyperbolic relaxation for
elliptic equations, some aspects have been explored
in~\cite{AlcBruDie02} in the context of ``gauge drivers'' for
numerical relativity.  In particular, \cite{AlcBruDie02} introduced
one of the most used gauge conditions for certain black hole
evolutions, the Gamma-driver for the shift vector, which employs a
hyperbolic equation related to the elliptic equation for a minimal
distortion shift. Also see~\cite{BalDauSei96} on gauge drivers, where
however only parabolic relaxation is considered.

The goal of the present paper is to develop hyperbolic relaxation
given by the prototype in (\ref{wavediffusion}) into a method to solve
a general class of second order, non-linear elliptic equations. The
basic observation that elliptic equations can be related to the
stationary end points of evolutions is well known. However, the
challenge is to develop a concrete formulation of a hyperbolic
relaxation method that can solve non-trivial problems.\footnote{ We
  are not aware of other work on hyperbolic relaxation introduced
  explicitly for the solution of non-trivial elliptic equations, but
  given the simplicity of the idea, it would be surprising if there is
  none. The authors welcome any pointers to existing literature.}  The
problem of immediate interest to us is defined by the constraint
equations of general relativity, which we solve as a system of
non-linear elliptic equations to obtain initial data for evolution in
numerical relativity. However, the formalism is quite independent of
this particular problem.

The main result of this paper is that hyperbolic relaxation can be
formulated for non-trivial equations, and numerical experiments for
some specific test problems were indeed successful. Specific examples
include the Poisson equation, the conformal-thin-sandwich equations,
scalar field initial data, and some simple configurations of single
and binary neutron stars.

Considering (\ref{wavediffusion}), let us collect some basic
observations here in order to introduce the main questions we want to
address. First of all, we have to address the well-posedness of the
hyperbolic PDEs. Given a self-adjoint, elliptic operator, the
hyperbolicity of equations of type (\ref{wavediffusion}) should be
clear. We demonstrate this below for a general class of
equations. There exists a rich theoretical background regarding
well-posedness and numerical stability for hyperbolic
PDEs~\cite{GusKreOli95,SarTig12,Hil13}, which helps to find relaxation
schemes that are well suited for numerical applications. However,
evolutions of hyperbolic PDEs are not trivial, so we should expect
that some elliptic problems are not amenable to hyperbolic relaxation
while others are, which is why we include a non-trivial set of test
problems below.

Second, in addition to the boundary conditions of the original
elliptic equation we have to choose boundary conditions for the
hyperbolic equations that are compatible with the asymptotic elliptic
problem This choice is not unique, but of great importance to obtain
successful evolutions. In particular, we consider maximally
dissipative boundary conditions.

Third, assuming feasibility and stability of hyperbolic relaxation, a
key question concerns the efficiency of the method. In both parabolic
and hyperbolic relaxation methods the time parameter is unrelated to
the elliptic equation, \ie the time evolution is of no interest as
long as the stationary state is reached efficiently. This is the basis
for different acceleration strategies. For hyperbolic relaxation,
there is a finite propagation speed, and in contrast to the diffusion
equation it is not clear how to by-pass that speed to accelerate the
method.

Beyond the intrinsic interest in a new method, we have to ask whether
hyperbolic relaxation, after some significant further development that
is beyond the scope of this paper, might become an interesting
alternative to the highly developed standard methods.

As it stands, there are pragmatic considerations that can make
hyperbolic relaxation methods interesting, in particular when solving
elliptic equations as part of a larger project. For example, elliptic
PDEs are often solved to provide initial data for evolution systems
that are subject to certain constraint equations, \eg the Maxwell
equations or the Einstein equations. However, the main work load is
the actual evolution of the data by integrating a hyperbolic PDE.  In
such a case the hyperbolic relaxation method does not have to compete
with optimized standard methods in terms of efficiency as long as
solving the elliptic equation is only a small part of the entire work
load.  On the other hand, a hyperbolic relaxation method may be easy
to implement using the existing infrastructure of a numerical
evolution code, avoiding the need for and the complications of an
external elliptic solver.  Using the same infrastructure also has the
advantage that interpolation errors can be avoided by using the same
grid discretization. Considering our research in numerical relativity,
a sophisticated infrastructure for evolutions is indeed available, but
we were looking for alternative elliptic solvers. Hence we implemented
hyperbolic relaxation in the pseudospectral hyperbolic evolution code
\texttt{bamps}~\cite{Bru11,HilWeyBru15}, which only required minor
modifications once the formalism itself was established.

The paper is structured as follows. In Sec.~\ref{Section:HypRelaxEqu}
we derive and motivate the evolution equations on which our relaxation
method is based and discuss some of its properties. In
Sec.~\ref{Section:Setup} we give details on the numerical
implementation and state the methods used in the \texttt{bamps} code.
In Secs.~\ref{Section:TestCases} and \ref{Section:Applications} we
apply our hyperbolic relaxation method to some test cases and to the
construction of initial data for numerical relativity. We conclude in
Sec.~\ref{Section:Conclusions}.

Throughout the paper we use the Einstein summation convention, \ie we
sum over indices that occur once as an upper index and once as a lower
index, \eg $s^i t_i = \sum_i s^i t_i$.  Latin letters $i, j, k, ...$
denote coordinate components and they are lowered and raised by an
arbitrary metric with positive signature.  An index $s$ denotes a
contraction with a vector $s_i$, in particular $\partial_s = s^i
\partial_i = s^i \frac{\partial}{\partial x^i}$.  Greek letter indices
$\alpha, \beta, \gamma$ denote components of a field and they are
lowered and raised by the Euclidean metric.  We also use Latin letters
$a, b, c$ to denote the spacetime components in general relativity and
which are lowered and raised by the spacetime metric $g_{ab}$.

\section{The Hyperbolic Relaxation Equations}\label{Section:HypRelaxEqu}
\subsection{Evolution System}\label{Subsection:EvolutionSystem}

In the following we present the principal ideas of the hyperbolic
relaxation method and derive the equations that follow for the
iteration scheme. In this paper we consider only systems of elliptic
equations given in second order form, \ie
\begin{equation}
\label{eq:ElliptEqu2ndOrder}
(\Lell \psi)_\alpha = {a(x^k)}\indices{^i^j^\beta_\alpha} \partial_i
\partial_j \psi_\beta + F_\alpha(x^k,\psi_\beta, \partial_i
\psi_\beta) = 0 \, ,
\end{equation}
where the $\psi_\alpha$ are the $N$ unknown solution variables and $F$
is a continuous function of the solution variables, their derivatives
and the $D$ coordinates $x^k$. We take~$a\indices{^i^j^\beta_\alpha}$
to be a smooth function of the coordinates and suppress this
dependence in the notation. In the following we consider
\emph{classically elliptic} systems~\cite{Dai04} only, \ie systems
with
\begin{equation}
\label{eq:DefClassEllipticity}
\det(a(x^k)\indices{^i^j^\beta_\alpha} s_i s_j) \neq 0 ~~~~\forall s
\in \mathbb{R}^{D}\setminus \{0\} \, ,
\end{equation}
where the determinant is understood to be taken on the indices
$\alpha$ and $\beta$.

Every elliptic system will be accompanied by a set of boundary
conditions on the variables $\psi_\alpha$ and we discuss their
treatment in section~\ref{Subsection:BoundaryConditions}. We can
reduce the second order elliptic system to first order by introducing
the \emph{reduction variables} $r_{i\alpha}$:
\begin{align}
\label{eq:ElliptEqu1stOrder_psi}
0 &= a\indices{^i^j^\beta_\alpha} \partial_i r_{j\beta} +
F_\alpha(\psi_\beta, r_{i\beta}) \, ,\\
\label{eq:ElliptEqu1stOrder_r}
0 &= \partial_i \psi_\alpha - r_{i\alpha} \, .
\end{align}
To solve the second order equation~\eqref{eq:ElliptEqu2ndOrder} one
could employ the Jacobi method, which can be motivated by evolving the
parabolic partial differential equation:
\begin{equation}
\label{eq:JacobiMethod}
\partial_t \psi_\alpha = (\Lell \psi)_\alpha \, ,
\end{equation}
where~$t$ is some parameter that plays the role of time.

For a classically elliptic system with constant coefficients the
Jacobi method can only converge if $a\indices{^j^i^\beta_\alpha}$ is
positive definite on the whole domain, \ie there exists an
$\epsilon>0$
\begin{equation}
\label{eq:Apositive}
a\indices{^j^i^\beta_\alpha} t_{j\beta} t\indices{_i^\alpha} \geq
\epsilon t^{i\alpha} t_{i\alpha} ~~~~\forall t \in \mathbb{R}^{D
  \times N}\setminus \{0\} \, ,
\end{equation}
which we assume in the rest of this paper. This condition corresponds
to the notion of \emph{strong ellipticity}, which defines an important
subclass of classically elliptic systems~\cite{Dai04}.  Note that we
have the freedom to multiply the elliptic equation with an invertible
matrix $d\indices{^\beta_\alpha}$, yielding $(\tilde \Lell \psi)_\alpha =
d\indices{^\beta_\alpha} (\Lell \psi)_\beta = 0$, which has the same
solutions as the original equation. This freedom allows us to
transform some systems into a strongly elliptic system, that
originally were not.

In analogy to the Jacobi method~\eqref{eq:JacobiMethod} we evolve
$\psi_\alpha$ by taking Eq.~\eqref{eq:ElliptEqu1stOrder_psi} as the
right-hand side, yielding
\begin{equation}
\label{eq:HypRelaxMethod_psi}
\partial_t \psi_\alpha = a\indices{^i^j^\beta_\alpha} \partial_i
r_{j\beta} + F_\alpha(x^i,\psi_\beta, r_{i\beta})
\end{equation}
and we proceed similarly with the equations for the reduction
variables $r_i$:
\begin{equation}
\label{eq:HypRelaxMethod_r}
\partial_t r_{i\alpha} = b\indices{^j_i^\beta_\alpha} ( \partial_j
\psi_\beta - r_{j\beta} ) \, ,
\end{equation}
where $b\indices{^j_i^\beta_\alpha}$ is arbitrary under the
requirement of positive definiteness, meaning in analogy to
Eq.~\eqref{eq:Apositive}
\begin{equation}
\label{eq:Bpositive}
b\indices{^j_i^\beta_\alpha} t_{j\beta} t^{i\alpha} > \epsilon
t_{i\alpha} t^{i\alpha} ~~~~\forall t \in \mathbb{R}^{D \times
  N}\setminus \{0\} \, .
\end{equation}
The system of Eqs.~\eqref{eq:HypRelaxMethod_psi}
and~\eqref{eq:HypRelaxMethod_r} forms a first order hyperbolic
differential equation which we refer to as the the \emph{hyperbolic
relaxation system}. Clearly the reduction constraint
Eq.~{\eqref{eq:ElliptEqu1stOrder_r} is not enforced at all times and
will indeed be violated during the relaxation process, however we
are only interested in the steady state, which fulfills the
reduction constraint, because Eq.~\eqref{eq:HypRelaxMethod_r}
constantly drives the reduction variable $r_i$ towards $\partial_i
\psi$. To see this, let us assume for arguments sake that
$\partial_t \partial_i \psi_\alpha = 0$. Then the solution for
$r_{i\alpha}$ has the form
\begin{equation}
\label{eq:rDecay}
r_{i\alpha}(t) = \sum_{l=1}^n e^{- \lambda_l t} \sum_{k_l=0}^{m_l}
{x^{k_l} h_{k_l i \alpha}} + \partial_i \psi_\alpha \, ,
\end{equation}
where $h$ is constant and the $\lambda_l$ are the $n$ eigenvalues
defined by the eigenvalue equation~$b\indices{^j_i^\beta_\alpha}
t\indices{_l^i^\alpha} = \lambda_l t\indices{_l^j^\beta}$ and $m_l$
depends on the geometric multiplicity of $\lambda_l$.  From the
positive definiteness of $b$ we know that all the eigenvalues have
positive real part and it follows immediately that $r_{i\alpha}$
approaches $\partial_i \psi_\alpha$ exponentially. We emphasize
however that in some cases, \eg if the elliptic system has no
solution, $\partial_i \psi_\alpha$ can grow faster than $r_{i \alpha}$
and thus the reduction constraints cannot be satisfied asymptotically
in time.

Assuming that the solution to the elliptic system exists and is unique
under the provided boundary conditions, then it is obvious that if the
hyperbolic relaxation system reaches a steady state, then we must have
the solution to the first order elliptic system
Eqs.~\eqref{eq:ElliptEqu1stOrder_psi}
and~\eqref{eq:ElliptEqu1stOrder_r}, and also of the original elliptic
equation \eqref{eq:ElliptEqu2ndOrder}.

The relationship between hyperbolic equations and parabolic diffusion
equations has already been investigated in some special
cases~\cite{HsiLiu92,Nis97,Wir14}.  In particular it can be shown that
for large times $t$ the solution of the hyperbolic equation
\begin{equation}
 \label{eq:HypRelax2ndOrder}
 \partial_t^2 \phi + \partial_t \phi - \Delta \phi = 0
\end{equation}
will tend towards the solution of the parabolic PDE
\begin{equation}
 \partial_t \phi - \Delta \phi = 0 \, .
\end{equation}
The hyperbolic equation can be cast in first order form by introducing
the reduction variables $\psi = \partial_t \phi + \phi$ and $r_i =
\partial_i \phi$, yielding the system
\begin{align}
 \label{eq:ParabolicDiff_RedPhi}
 \partial_t \phi &= \psi - \phi \, , \\
 \label{eq:ParabolicDiff_RedPsi}
 \partial_t \psi &= \delta^{ij} \partial_i r_j \, , \\
 \label{eq:ParabolicDiff_Redr}
 \partial_t r_i &= \partial_i \psi - r_i \, .
\end{align}
Clearly the first of these equations is an ordinary differential
equation that has no direct counterpart in our hyperbolic relaxation
system, however it is directly evident that $\phi$ will tend towards
$\psi$ exponentially. Thus it is plausible that for large $t$ the
variable $\psi$ of our hyperbolic relaxation system will behave
similarly to that of the Jacobi-type relaxation
equation~\eqref{eq:JacobiMethod}.

\subsection{Residual Evolution}\label{Subsection:ResEvo}

The residuals of the first order system,
Eqs.~\eqref{eq:ElliptEqu1stOrder_psi},\eqref{eq:ElliptEqu1stOrder_r},
are given by
\begin{align}
\label{eq:ResidualR}
R_\alpha &= a\indices{^i^j^\beta_\alpha} \partial_i r_{j\beta} +
F_\alpha(x^i,\psi_\beta, r_{i\beta}) \, , \\
\label{eq:ResidualRi}
R_{i\alpha} &= b\indices{^j_i^\beta_\alpha} ( \partial_j \psi_\beta -
r_{j\beta} ) \, .
\end{align}
A simple calculation shows that the residuals will evolve according to
\begin{align}
\label{eq:ResidualR_evo}
\partial_t R_\alpha &= a\indices{^i^j^\beta_\alpha} \partial_i
R_{j\beta} + \frac{\partial F_\alpha}{\partial \psi_\beta} R_\beta +
\frac{\partial F_\alpha}{\partial r_{i \beta}} R_{i \beta} \, , \\
\label{eq:ResidualRi_evo}
\partial_t R_{i\alpha} &= b\indices{^j_i^\beta_\alpha} ( \partial_j R_\beta - R_{j\beta} ) \, .
\end{align}
For a working relaxation scheme, we want the residual evolution system
to be stable, \ie the first order residuals should converge to zero
for $t \rightarrow \infty$, for residuals that are sufficiently close
to zero.  Systems of this type and stability conditions are discussed
in detail in~\cite{KreOrtReu98} and~\cite{KreLor98}. It is not
possible for us to give general results on the stability of the
hyperbolic relaxation scheme, as the multitude of possible systems is
too large to be covered in a closed form, especially for elliptic
systems with more than one variable.  A stability analysis must
therefore be done individually for the concrete problem.

\subsection{Mode Analysis}\label{Subsection:ModeAna}

To shed some light on the behavior of solutions to the hyperbolic
relaxation equation \eqref{eq:HypRelax2ndOrder}, we perform a simple
mode analysis, ignoring the issue of boundary conditions. Introduce
the plane-wave ansatz
\begin{equation}
  \phi_{pw}(t,x) = e^{i(k x - \omega t)},
\end{equation}
where $k$ and $\omega$ are constants. The wavenumber $k$ is a real
number related to the wave length, $k = 2\pi/\lambda$, while $\omega$
may be a complex number.  Inserted into the hyperbolic relaxation
equation \eqref{eq:HypRelax2ndOrder}, we obtain
\begin{align}
  \omega^2 + i \omega &= k^2,\\
  \omega_\pm(k) &= - \frac{1}{2} (i \pm \sqrt{4k^2 - 1}).
\end{align}
Recall that for the wave equation $\omega_\pm(k) = \pm k$, while for
the heat equation $\omega(k) = -i k^2$. For hyperbolic relaxation,
there is a further case distinction for the sign under the square root
$\sqrt{4k^2 - 1}$.

For sufficiently large wavenumber,
\begin{equation} \phi_{pw} =
  e^{-\frac{1}{2}t} e^{i(k x \pm \frac{1}{2}\sqrt{4k^2 - 1}\,t)},
  \quad k \geq \frac{1}{2},
\end{equation}
which is a damped wave with phase velocity
$v(k)=\sqrt{1-\frac{1}{4k^2}}$.  The damping is independent of $k$ (as
opposed to the heat equation with $e^{-k^2 t}$). The phase velocity
approaches $v=1$ for large $k$, but for $k$ approaching the critical
value $\frac{1}{2}$ from above the phase velocity tends towards $v=0$.

For sufficiently small wavenumber,
\begin{equation}
   \phi_{pw} = e^{-\frac{1}{2}(1\pm\sqrt{1-4k^2})t} e^{i k x}, \quad 0
   \leq k \leq \frac{1}{2},
\end{equation}
which is a non-moving wave profile $e^{i k x}$ times a $k$-dependent
damping factor.  For $k = \frac{1}{2}$, the damping is
$e^{-\frac{1}{2}t}$, while for $k$ equal to zero there are two cases,
$-\frac{1}{2}(1\pm\sqrt{1-4k^2})t = 0$ or $-t$. For small $k$, the
worse (more weakly) damped case is $-\frac{1}{2}(1\pm\sqrt{1-4k^2})t
\approx - k^2 t$, which is the same damping as for the basic heat
equation.

\begin{figure}[tb]
    \includegraphics[width=0.4\textwidth]{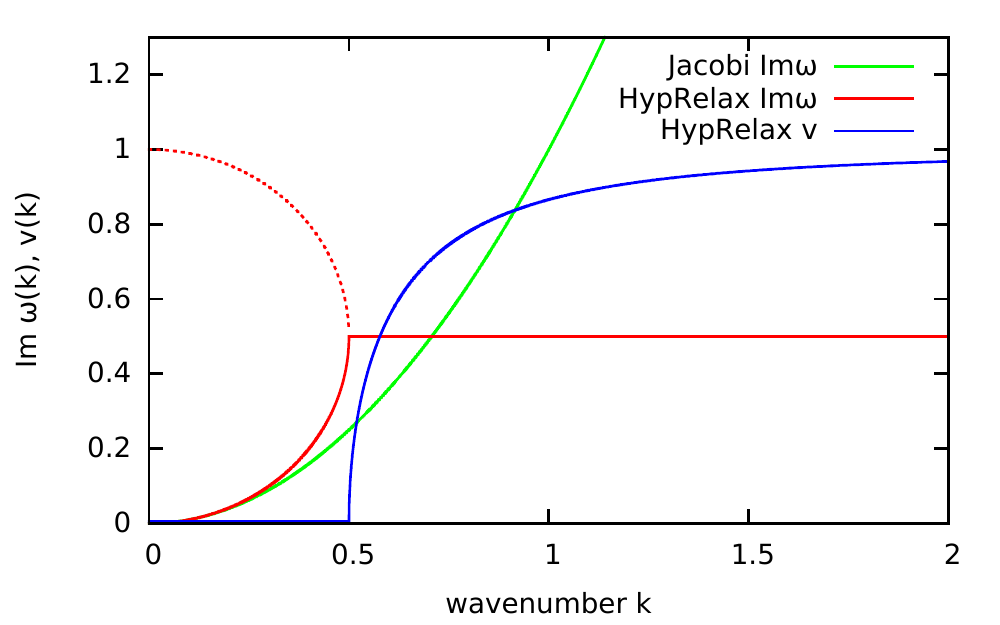}    
    \caption{
      \label{fig:ModesOmegaV}
      Damping and propagation speed of basic hyperbolic relaxation
      compared to parabolic relaxation. There is a transition at
      $k_{crit}=\frac{1}{2}$, which can be moved to lower values by
      introducing additional parameters. The overall damping is
      determined by the slowest damping rate.  }
\end{figure}

Summarizing, the plane-wave mode analysis suggests that solutions to
the hyperbolic relaxation equation exhibit a mixture of relaxation and
wave propagation phenomena, see Fig.~\ref{fig:ModesOmegaV}.  For wave
numbers larger than a critical value, $k \geq k_{crit}$ with $k_{crit}
= \frac{1}{2}$, there is wave propagation with simultaneous damping.
Noteworthy is that the damping is independent of $k$,
$e^{-\frac{1}{2}t}$. This is a promising feature compared to parabolic
relaxation with $e^{-k^2 t}$ for intermediate values of $k$.  For
large values of $k$ parabolic relaxation has much stronger damping,
but the overall convergence rate is dominated by small $k$.  For
hyperbolic relaxation, there is no wave propagation for $k \leq
k_{crit}$, but the damping persists. Interestingly, the damping factor
asymptotes towards $e^{-k^2 t}$ for $k\rightarrow0$, and is never
worse than parabolic relaxation for small $k$.

The existence of a transition at a specific length scale
$\lambda_{crit}=\frac{2\pi}{k_{crit}} = 4\pi$ signals that our choice
of hyperbolic relaxation equation has fixed a scale. Let us
generalize~\eqref{eq:HypRelax2ndOrder} to
\begin{equation}
  \epsilon \partial_t^2 \phi + \eta \partial_t \phi = \Delta \phi,
\label{hyprelax_epseta}
\end{equation}
where $\epsilon$ and $\eta$ are real, non-negative constants. A third
constant in front of $\Delta \phi$, the velocity squared $v^2$ in the
basic wave equation, has been rescaled to one without loss of
generality. With $\epsilon=\eta=1$ as in \eqref{eq:HypRelax2ndOrder}
we fix the unit of time to be dimensionless (unity since $[T^2]=[T]$).
If we mimic (\ref{eq:HypRelaxMethod_psi}) and
(\ref{eq:HypRelaxMethod_r}) by
\begin{equation}
\partial_t \psi = \partial_j r^j, \quad \partial_t r_i = b (\partial_i
\psi - r_i),
\label{hyprelax_b}
\end{equation}
with a real constant $b$, then we obtain \eqref{hyprelax_epseta} with
$\epsilon = 1/b$, $\eta=1$, and
\begin{equation}
\omega_\pm(k) = - \frac{1}{2} (i b \pm \sqrt{(4k^2 - b)b}).
\end{equation}
Alternatively, we can take our motivation from the gauge driver
construction \cite{AlcBruDie02} and set $\epsilon = 1$, $\eta$ an
arbitrary non-negative constant, and obtain
\begin{equation}
\omega_\pm(k) = - \frac{1}{2} (i \eta \pm \sqrt{4k^2 - \eta^2}).
\end{equation}
We also considered $\eps=\eta^2$ and arbitrary $\eta$ for uniform
scaling of time.

The bottom line is that the coefficients in \eqref{hyprelax_epseta}
allow us to adjust the size of the critical length scale. For example,
for $\epsilon = 1$ and $\eta$ arbitrary the critical parameters change
to $k_{crit} = \frac{\eta}{2}$, while for $\epsilon = 1/b$, $\eta=1$
we find $k_{crit} = \frac{\sqrt{b}}{2}$. The damping for large $k$
becomes $e^{-\frac{\eta}{2}}$ and $e^{-\frac{b}{2}}$, respectively.

To avoid small $k$ that drop below (or too far below) $k_{crit}$, we
can adjust $b$ or $\eta$ such that the length scale of $k$ corresponds
to the physical size $L$ of the domain, say $\lambda_{crit} = 2
L$. This will slow down the convergence for large $k$, but will also
avoid the severe slow down when the damping approaches that of
parabolic relaxation.

\subsection{Hyperbolicity Analysis}\label{Subsection:HyperbolicityAnalysis}

In this section we introduce a notation for inverse tensors. These
have to be understood as the inverse of matrices with respect to the
field indices. For example the inverse of the tensor
$c\indices{^\beta_\alpha}$ is $(c^{-1})\indices{^\alpha_\gamma}$ and
we have
\begin{equation}
 c\indices{^\beta_\alpha} (c^{-1})\indices{^\alpha_\gamma} =
 \delta^\beta_\gamma \, .
\end{equation}

For the hyperbolicity analysis we start with writing the hyperbolic
relaxation system in matrix form
\begin{equation}
\label{eq:HypRelaxSystem}
\partial_t {\bf u} = {\bf P}^k \partial_k {\bf u} + h(x^i,{\bf u})
\end{equation}
with
\begin{equation}
\label{eq:HypRelPrincMatrix}
{\bf P}^k =
\begin{pmatrix}
  0                              & a\indices{^k^i^\beta_\alpha} \\
  b\indices{^k_i^\beta_\alpha} & 0
\end{pmatrix} \, ,
~~~~~~
{\bf u} = 
\begin{pmatrix}
  \psi_\alpha \\
  r_{i\alpha}
\end{pmatrix} \, .
\end{equation}
The principal symbol of this system is then given by
\begin{equation}
\label{eq:HypRelaxPrincSymbol}
{\bf P}^s = {\bf P}^k s_k = 
\begin{pmatrix}
  0   & a\indices{^s^i^\beta_\alpha} \\
  b\indices{^s_i^\beta_\alpha} & 0 
\end{pmatrix} \, ,
\end{equation}
where $s_i$ is an arbitrary unit vector, $s^i s_i = 1$. Suppose
$c\indices{^\beta_\alpha} = a\indices{^s^i^\epsilon_\alpha}
b\indices{^s_i^\beta_\epsilon}$ has a complete set of eigenvectors
$w\indices{_\gamma^\alpha}$ with $c\indices{^\beta_\alpha}
w\indices{_\gamma^\alpha} = \sigma^\alpha_\gamma
w\indices{_\alpha^\beta} $, where $\sigma^\alpha_\gamma$ is
diagonal. If furthermore all the eigenvalues, \ie the diagonal
elements of $\sigma^\alpha_\gamma$, are positive then $P^s$ has the
following left eigenvectors
\begin{align}
\label{eq:HypRelaxEigenvectors0}
      {\bf v}^0_{j \gamma} &= \left ( 0 , \delta^\alpha_\gamma
      \delta^i_j - b\indices{^s_j^\epsilon_\gamma} (
            {b\indices{^s_s}}^{-1} )\indices{^\alpha_\epsilon} s^i
            \right ) \, ,\\
\label{eq:HypRelaxEigenvectors+-}
      {\bf v}^{\pm}_\gamma &= \left ( w\indices{_\gamma^\alpha}, \pm
      (\rho^{-1})^\delta_\gamma w\indices{_\delta^\epsilon}
      a\indices{^s^i^\alpha_\epsilon} \right ) \, ,
\end{align}
where $\rho^\beta_\gamma$ is the root of $\sigma^\beta_\gamma$, \ie it
is a positive diagonal tensor with $\rho^\alpha_\gamma
\rho^\beta_\alpha = \sigma^\beta_\gamma$.  Note that of the
eigenvectors ${\bf v}^0_{j\gamma}$ only $(D-1)N$ are linearly
independent, while the ${\bf v}^{\pm}_\gamma$ are $2 N$ independent
vectors. If there exists a constant~$K$, independent of~$s^i$, such
that $\|{\bf V}\|_2 + \|{\bf V}^{-1}\|_2 < K$, where ${\bf V}$ is an,
in general $s^i$-dependent, square matrix constructed from a linearly independent 
set of the eigenvectors ${\bf v}$, then the system is strongly
hyperbolic~\cite{GusKreOli95,Hil13,SarTig12}. The characteristic
variables $\hat u$ and their characteristic speeds $\lambda$ are thus
\begin{align}
\label{eq:HypRelaxCharacteristics0}
\hat u^0_{j\gamma} &= r_{j\gamma} - b\indices{^s_j^\epsilon_\gamma} (
     {b\indices{^s_s}}^{-1} )\indices{^\alpha_\epsilon} s^i
     r_{i\alpha} \, , ~~~~~~~~ \lambda^0_{i\gamma} = 0 \, , \\
\label{eq:HypRelaxCharacteristics+-}
\hat u^{\pm}_\gamma &= w\indices{_\gamma^\alpha} \psi_\alpha \pm
(\rho^{-1})^\delta_\gamma w\indices{_\delta^\epsilon}
a\indices{^s^i^\alpha_\epsilon} r_{i\alpha} \, , ~~~
\lambda_\gamma^\pm = \pm \rho^\alpha_\gamma e_\alpha \, ,
\end{align}
where $e_\alpha$ denote the Cartesian basis vectors. Note that of the
characteristic variables $\hat u^0_{j\gamma}$ only $(D-1)N$ are
linearly independent, while the $\hat u^{\pm}_\gamma$ are $2 N$
independent vectors.  From this we can recover the evolved variables
in terms of the characteristics:
\begin{align}
\psi_\alpha &= \frac{1}{2} ( w^{-1} )\indices{_\alpha^\gamma} (\hat
u^+_\gamma + \hat u^-_\gamma) \, , \\
r_{i\alpha} &= \hat
u^0_{i\alpha} + b\indices{^s_i^\epsilon_\alpha} ( c^{-1}
)\indices{^\beta_\epsilon} \nonumber \\
& \qquad \qquad \cdot \left (
(w^{-1})\indices{_\beta^\delta} \rho^\gamma_\delta \frac{\hat
  u^+_\gamma - \hat u^-_\gamma}{2} - a\indices{^s^j^\gamma_\beta} \hat
u^0_{j \gamma} \right ) \, .
\end{align}

We can use the freedom in the choice of $b\indices{^j_i^\beta_\alpha}$
to impose certain properties on the hyperbolic relaxation system. In
the following we briefly discuss three interesting choices, that
fulfill the restrictions we have set for $b$.

\emph{1. $b$ is the identity.} A very easy and natural choice is
$b\indices{^j_i^\beta_\alpha} = \delta^j_i \delta^\beta_\alpha$. With
this choice we have $c\indices{^\beta_\alpha} =
a\indices{^s^s^\beta_\alpha}$, which has only eigenvalues with
positive real part due to~\eqref{eq:Apositive}. The imaginary part
however can be non-vanishing.  If however
$a\indices{^s^s^\alpha_\gamma} = a\indices{^s^s_\gamma^\alpha}$, then
$c$ is guaranteed to have a complete set of eigenvectors with purely
real eigenvalues and thus system is strongly hyperbolic.  If we have
$a\indices{^i^j^\alpha_\gamma} = a\indices{^i^j_\gamma^\alpha}$, then
the system is even symmetric hyperbolic with symmetrizer:
\begin{equation}
\label{eq:BIdentitySymmetrizer}
{\bf H} = 
\begin{pmatrix}
  \delta^\alpha_\gamma & 0 \\
  0 & a\indices{^i^j^\alpha_\gamma}
  \delta_{j l}
\end{pmatrix} \, .
\end{equation}

\emph{2. $b$ is the transpose of $a$.} We can also make the system
trivially symmetric hyperbolic by choosing
$b\indices{^j_i^\beta_\alpha} = a\indices{^j_i_\alpha^\beta}$.  The
principal symbol of this system is symmetric and thus the system is
symmetric hyperbolic.

\emph{3. $b$ is the inverse of $a$.} We can choose $b$ to be the
inverse of $a$ in the sense that $b$ fulfills
$a\indices{_k^i^\alpha_\gamma} b\indices{^j_i^\beta_\alpha} =
\delta_k^{j} \delta^\beta_\gamma$. This choice is particularly
interesting because we then have $c\indices{^\beta_\alpha} =
\delta^\beta_\alpha$ and thus all the non-zero characteristic speeds
have values $\pm 1$. Furthermore the eigenvectors of $c$ become
trivial: $w\indices{_\gamma^\alpha} = \delta^\alpha_\gamma$.  A
symmetrizer for this system is
\begin{equation}
\label{eq:BInverseSymmetrizer}
{\bf H} =
\begin{pmatrix}
  \delta^\alpha_\gamma & 0 \\
  0 & a\indices{_m^i^\alpha_\omega} a\indices{^m_l_\gamma^\omega} 
\end{pmatrix} \, .
\end{equation}
Since all traveling characteristic variables have the same speeds, we
consider this the best choice for $b\indices{^j_i^\beta_\alpha}$.  A
straightforward generalization of this choice allows
$b\indices{^j_i^\beta_\alpha}$ to be scaled by a constant factor,
which will also uniformly scale the characteristic speeds.

\subsection{Boundary Conditions}
\label{Subsection:BoundaryConditions}

The basic idea to impose boundary conditions in our method is to
modify the right hand side of the hyperbolic relaxation system. The
outward pointing unit normal covector~$s_i$ to the boundary surface is
naturally defined by taking the gradient of a scalar field which is
increasing across, but constant in the boundary, and then normalizing
this gradient to unit magnitude using our arbitrary but fixed
metric. This metric is subsequently used to raise the index and form
the outward pointing vector~$s^i$. We restrict our attention to
strongly hyperbolic systems, for which a regular ($s^i$-dependent)
similarity transformation matrix~${\bf T}_s$ exists which transforms
between our evolved variables~${\bf u}$ and a linearly independent set
of the characteristic variables ${\bf \hat u}$ given in
Eq.~\eqref{eq:HypRelaxCharacteristics0}~and~\eqref{eq:HypRelaxCharacteristics+-}
\begin{equation}
\label{eq:characteristicTrafo}
{ \bf u} = {\bf T}_s {\bf \hat u} \, .
\end{equation}
We can then decompose our evolution
equations~\eqref{eq:HypRelaxSystem} as
\begin{equation}
\label{eq:HypRelaxSystemSplit}
\partial_t { \bf u} = {\bf P}^s \partial_s { \bf u} + {\bf P}^k q_k^i
\partial_i { \bf u} + h(x^i,{ \bf u}) \, ,
\end{equation}
where $q_k^i = \delta_k^i - s_k s^i$ is the projector onto the
boundary surface. We now multiply by ${\bf T}_s^{-1}$ and obtain
\begin{align}
\label{eq:HypRelaxSystemSplitTrans}
{\rm d}_t {\bf \hat u} =& {\bf T}_s^{-1} {\bf P}^s \partial_s { \bf u}
+ {\bf T}_s^{-1} ({\bf P}^k q_k^i \partial_i { \bf u} + h(x^i,{ \bf
  u}) ) \\
=& {\bf T}_s^{-1} {\bf P}^s T_s {\bf T}_s^{-1} \partial_s {
  \bf u} + T_s^{-1} ({\bf P}^k q_k^i \partial_i { \bf u} + h(x^i,{ \bf
  u}) ) \\
=& {\bf \Lambda}^s {\rm d}_s \hat { \bf u} + {\bf T}_s^{-1}
({\bf P}^k q_k^i \partial_i { \bf u} + h(x^i,{ \bf u}) ) \, .
\end{align}
Here the straight derivative symbol ${\rm d}$ denotes that the
transformation matrix stands outside of the derivative, \ie ${\rm d}_i
\hat { \bf u} \equiv {\bf T}_s^{-1} \partial_i { \bf u}$, and ${\bf
  \Lambda}^s$ is a diagonal matrix containing the characteristic
speeds.  We can now impose boundary conditions on the incoming
variables, \ie those with positive characteristic speeds, by modifying
their right hand sides.  After the right hand sides have been modified
we transform the system back by multiplying with ${\bf T}_s$.

\subsubsection{Penalty Method}
\label{Subsubsection:BoundaryConditionsPenalty}

In the penalty method~\cite{Hes00,HesGotGot07,TayKidTeu10} the
boundary conditions are \emph{weakly imposed} by modifying the right
hand sides of the incoming characteristic variables $\hat u^+_\gamma$
in the following way
\begin{equation}
  \label{eq:PenaltyMethod}
{\rm d}_t \hat u^+_\gamma \hat = D_t \hat u^+_\gamma + p (\hat u^{+
  \rm BC}_\gamma - \hat u^+_\gamma) \, ,
\end{equation}
where $p$ is the penalty parameter, $u^{+ \rm BC}_\gamma$ is some
given boundary data that we want $\hat u^+_\gamma$ to approach, $D_t
\hat u^+_\gamma$ is the unmodified right hand side and $\hat =$
denotes equality at the boundary. The penalty parameter can not be
chosen arbitrarily and we refer the reader to~\cite{HilWeyBru15} for a
detailed derivation of the penalty parameters used in \texttt{bamps}.

\subsubsection{Maximally Dissipative Boundary Conditions}
\label{Subsubsection:BoundaryConditionsMaxDiss} 

Maximally dissipative boundary
conditions~\cite{GusKreOli95,GunGar05,SarTig12,Hil13} will allow us to
set boundary conditions of the form
\begin{equation}
\label{eq:BCpsiDeriv}
s^i \partial_i \psi_\alpha|_{\partial \Omega} =
\phi_\alpha(\psi_\beta, \partial_i \psi_\beta, q^i_j \partial_i
\partial_k \psi_\beta ) \, ,
\end{equation}
where $\phi$ is a function that is allowed to depend on the
coordinates $x^i$, the fields $\psi_\alpha$, their derivatives and the
transverse projections ($q^i_j = \delta^i_j - s^i s_j$) of their
second derivatives.  For brevity we suppress dependence on all the
arguments in the following.

Maximally dissipative boundary conditions are imposed by requiring
\begin{equation}
\label{eq:BCMaxDissPsi}
(\rho^{-1})^\beta_\gamma w\indices{_\beta^\alpha} \partial_t
\psi_\alpha + w\indices{_\gamma^\alpha} \partial_s \psi_\alpha =
w\indices{_\gamma^\alpha} \phi_\alpha \, .
\end{equation}
This boundary condition is actually be different from
Eq.~\eqref{eq:BCpsiDeriv} during the relaxation process. However,
again we are only interested in the steady state at the end of the
evolution, where we have $\partial_t \psi_\alpha = 0$ and thus the
correct boundary condition will be imposed. For numerical stability
the functions $\phi_\alpha$ must not depend on normal derivatives of
the evolved variables. Therefore in~\eqref{eq:BCpsiDeriv} in the
arguments of $\phi_\alpha$ we have to make the replacements
$\partial_i \psi_\beta \rightarrow r_{i \beta}$ and $q^i_j \partial_i
\partial_k \psi_\beta \rightarrow q^i_j \partial_i r_{k \beta}$.  For
the normal derivatives of the incoming characteristic we obtain the
relation
\begin{align}
{\mathrm d}_s \hat u^+_\gamma &= w\indices{_\gamma^\alpha} \partial_s
\psi_\alpha + (\rho^{-1})^\delta_\gamma w\indices{_\delta^\epsilon}
a\indices{^s^i^\alpha_\epsilon} \partial_s r_{i\alpha} \\
\label{eq:BCMaxDissU}
    &= w\indices{_\gamma^\alpha} \phi_\alpha -
(\rho^{-1})^\delta_\gamma w\indices{_\delta^\alpha} \left ( \partial_t
\psi_\alpha - a\indices{^s^j^\epsilon_\alpha} \partial_s r_{j
  \epsilon} \right ) \, ,
\end{align}
where in the actual implementation $\partial_t \psi_\alpha$ is to be
replaced by Eq.~\eqref{eq:HypRelaxMethod_psi}.  This equation is now
used to impose the boundary condition by replacing the ${\mathrm d}_s
\hat u^+_\gamma$terms in Eq.~\eqref{eq:HypRelaxSystemSplitTrans},
yielding the modified right hand side
\begin{align}
{\mathrm d}_t \hat u^+_\gamma \hat =& D_t \hat u^+_\gamma \nonumber\\
& - \rho^\beta_\gamma ( w\indices{_\beta^\alpha} \partial_s
\psi_\alpha + (\rho^{-1})^\delta_\beta w\indices{_\delta^\alpha}
\partial_t \psi_\alpha - w\indices{_\beta^\alpha} \phi_\alpha ) \, .
   \label{eq:BCMaxDissModRHS}
\end{align}
With the general expression at hand, we now discuss the implementation
of typical boundary conditions.

\emph{1. Dirichlet conditions.} Dirichlet conditions are of the form
$\psi_\alpha |_{\partial \Omega} = g_\alpha$, where the $g_\alpha$ are
some function defined on the domain boundary $\partial \Omega$. To
implement such a boundary condition $\phi_\alpha$ has to take the form
\begin{equation}
\label{eq:BCDirichletPhi}
\phi_\alpha = s^i r_{i \alpha} + e\indices{^\beta_\alpha} ( g_\beta - \psi_\beta) \, ,
\end{equation}
where $e$ is positive definite, \ie $e\indices{^\beta_\alpha} t_\beta
t^\alpha > 0$.  In the steady state we have $\partial_i \psi_\alpha =
r_{i \alpha}$ and thus Eq.~\eqref{eq:BCMaxDissPsi} becomes $0 =
e\indices{^\beta_\alpha} ( g_\beta - \psi_\beta)$, which is only
fulfilled for the requested boundary condition.  The positive
definiteness of $e$ is important to guarantee stability at the
boundary. Suppose we have $\partial_i \psi_\alpha = r_{i \alpha}$
fixed, then Eq.~\eqref{eq:BCMaxDissPsi} has the form
\begin{equation}
\label{eq:BCDirichletStability}
  (\rho^{-1})^\beta_\gamma w\indices{_\beta^\alpha} \partial_t
\psi_\alpha = w\indices{_\gamma^\alpha} e\indices{^\beta_\alpha} (
g_\beta - \psi_\beta) \, ,
\end{equation}
which would have solutions not asymptoting to $g_\alpha$ if $e$ was
not positive definite.  Besides positive definiteness there are no
further restrictions apparent on $e$ and therefore, it can be chosen
to be the identity $e\indices{^\beta_\alpha} = \delta^\beta_\alpha$,
which we use in our applications.

\emph{2. Neumann conditions.} Neumann boundary conditions are of the
form $s^i \partial_i \psi_\alpha|_{\partial \Omega} = g_\alpha$.
Their implementation in our method is straightforward; one just has to
take $\phi_\alpha = g_\alpha$.

\emph{3. Robin conditions.} Robin boundary conditions are mixture of
Dirichlet and Neumann boundary conditions and can be written as $s^i
\partial_i \psi_\alpha|_{\partial \Omega} = g_\alpha + f^\beta_\alpha
\psi_\beta$, where the $f^\beta_\alpha$ are functions defined on the
domain boundary.  Their implementation is also straight forward
choosing $\phi_\alpha = g_\alpha + f^\beta_\alpha \psi_\beta$.

\section{Numerical Setup}\label{Section:Setup}
\subsection{Grid Setup}\label{Subsection:GridSetup}

We employ the pseudospectral hyperbolic evolution code \texttt{bamps}
and refer the reader to~\cite{HilWeyBru15}, where the grid setup is
explained in detail.  Here we only give a short summary of the basic
grid setup and numerical method.  Our grid consists of different
coordinate patches, a cube patch in the center, transition shell
patches and outer shell patches. On each patch we have a mapping
between local Cartesian coordinates to global Cartesian coordinates,
where on shell patches we employ the ``cubed sphere''
construction~\cite{RonIacPao96}.  The patches themselves can consist
of smaller subpatches, which are the smallest units we use for our
parallelization scheme. On each subpatch we approximate the fields by
a Chebyshev pseudospectral method, \ie the subpatches are discretized
in every direction by the Gauss-Lobatto collocation points. It is then
possible to reconstruct the Chebyshev coefficients from the fields
values at the collocation points.

The \texttt{bamps} code is adapted to evolutions in three dimensions.
For axisymmetric and spherically symmetric problems we use the cartoon
method to reduce the computational domain to two or one dimensions
respectively~\cite{HilWeyBru15}.

\subsection{Integration Method}\label{Subsection:Integrator}

The time integration for relaxation methods does not require a high
order of error convergence, since we are only interested in the steady
state at the end of the evolution.  More important are the efficiency
and stability of the integration algorithm.  For the time integration
we use the method of lines.  It is known that for linear hyperbolic
equations the simple forward Euler-method and also explicit
second-order Runge-Kutta methods are unstable~\cite{GusKreOli95} and
thus are not suited for the integration of the hyperbolic relaxation
equations.

In the applications presented in this paper we employ the popular
fourth-order Runge-Kutta scheme (RK4), which is stable for hyperbolic
equations.  This method needs four evaluations of the right-hand side
per time step, which appears to be not very efficient. After all, we
do not need a very accurate integrator, since we are only interested
in approaching the stationary state.  Therefore it is worthwhile to
investigate other classes of integrators, \eg multistep methods like
the third- or fourth-order Adams schemes~\cite{Boy01}, which
effectively only require one or two evaluations per time step and are
usually also stable for hyperbolic PDEs. Some simple experiments with
the Poisson equation indicated, however, that RK4 is more efficient
than RK3 or a fourth order Adams scheme since RK4 allows comparatively
large time steps.

Contrary to what is described in~\cite{HilWeyBru15}, we neither use
nor need filtering to assure stability in the hyperbolic relaxation
method, since the system usually tends towards a stable static or
stationary solution automatically.

In our code we have two types of boundaries.  On the one hand we have
boundaries between different subpatches, and on the other hand the
boundaries of our computational domain, in particular the outer
boundaries.  To treat boundaries between subpatches we employ the
penalty method described in
Sec.~\ref{Subsubsection:BoundaryConditionsPenalty} setting the
boundary data to be the outgoing characteristic of the neighboring
subpatch.

Treating the outer boundary with this method we have to provide a
function $g_\gamma$ equaling $u^{+ \rm BC}_\gamma$ at the boundary, \ie
\begin{equation}
\label{eq:PenaltyOuterBC}
\hat u^{+ \rm BC}_\gamma = w\indices{_\gamma^\alpha} \psi_\alpha 
+ (\rho^{-1})^\delta_\gamma w\indices{_\delta^\epsilon}
a\indices{^s^i^\alpha_\epsilon} r_{i\alpha}  = g_\gamma\, .
\end{equation}
Such a boundary condition however is very unusual in practice for
elliptic equations and thus the penalty method is not suited for the
treatment of our outer boundaries. Instead we only use the maximally
dissipative boundary conditions described in
Sec.~\ref{Subsubsection:BoundaryConditionsMaxDiss}.

\subsection{Initial Guesses}\label{Subsection:InitialGuesses}

To start the hyperbolic relaxation one has to provide an initial guess
to the solution. A suitable initial guess will always depend on the
specific form of the problem, in particular it should be chosen such
that in the course of the relaxation the variables do not have to
cross any points where the equations (\eg terms in the non-principal
part) become singular.  In our tests we found the solver to be
particularly well behaved when starting with a guess that is
stationary in the interior, but not at the boundary.  The whole
solution then starts to relax from the boundary to the inside.  For
example in our applications to numerical relativity initial data,
taking the flat metric everywhere lead to stable relaxations, which
demonstrates a remarkably high robustness that can be achieved by the
method.  For the reduction variables we simply take the numerical
derivative of the initial guess, \ie $^{(\rm ini)}r_{i \alpha} =
\partial_i {^{(\rm ini)}\psi_\alpha}$

\subsection{Refinement Strategy}\label{Subsection:RefinementStrategy}

To speed up the relaxation process we employ a simple scheme of
successive refinement.  It can be assumed that the right-hand side of
the solution variables $\partial_t \psi_\alpha$,
Eq.~\eqref{eq:HypRelaxMethod_psi}, is a good approximation to the
residual of the elliptic equation $(L\psi)_\alpha$,
Eq.~\eqref{eq:ElliptEqu2ndOrder}.  This however is only true until a
discretization limit is reached below which the norm of the residual
is no longer decreasing. The norm of the $\partial_t \psi_\alpha$ will
typically continue decreasing until machine precision is reached. This
makes it possible to construct an indicator signaling when the
discretization limit is reached and thus relaxation should be
continued on a higher resolution grid.  In particular we choose the
following criterion,
\begin{equation}
 \label{eq:RefineCriterion}
 \int \sum_{\alpha = 1}^N | \partial_t \psi_\alpha | dV
 < c \int \sum_{\alpha = 1}^N | (L\psi)_\alpha | dV \, ,
\end{equation}
where $c$ is some constant smaller than one. For our applications we
found $c = 0.1$ to be a choice working reasonably well. We note
however that depending on the specific problem also smaller values
might be beneficial.  Additionally we increase the resolution when the
error of the elliptic equation reaches machine precision, \ie when the
norm of $(L\psi)_\alpha$ is smaller than $10^{-13}$ times the number
of grid points.

We start with relaxing the system on the coarsest grid and check every
1000 relaxation steps whether to proceed relaxation on a finer grid
based on whether one of the two criteria mentioned above is fulfilled.
The final resolution can be determined by an error bound on the
residual of the elliptic equation, or by some predetermined
resolution, which may be required for the evolution of the data.  For
the refinement we increase the resolution on every subpatch by two
collocation points in every direction, which is equivalent to adding
two Chebyshev modes in every direction.  We interpolate the coarse
steady state solution to the new subpatches and repeat the procedure
until we arrive at the desired resolution.  We also find it advisable
to use the interpolated values for the reduction variables instead of
taking the numerical derivative of the solution variables, since the
latter introduces new errors, which costs some extra effort to damp.

\section{Application to Test Cases}\label{Section:TestCases}
\subsection{Poisson Equation -- Finite Differencing }
\label{Subsection:PoissonFD}

To provide a reference point independent of the specific
pseudospectral methods of \texttt{bamps}, we first discuss a minimal
implementation using a finite difference method to solve the Poisson
equation.  We consider the hyperbolic relaxation equation
\begin{equation}
\partial_t^2 \phi + \eta \partial_t \phi = \Delta \phi - \rho,
\end{equation}
which we implement as a first order in time, second order in space system,
\begin{align}
  \partial_t \phi &= \pi - \eta \phi,\\
  \partial_t \pi &= \Delta
    \phi - \rho.
\end{align}
We consider the fully first order version of this system in
Sec.~\ref{Subsection:Poisson}. At the boundaries we use asymptotic
Dirichlet conditions analogous to \eqref{eq:BCDirichletPhi},
$\partial_t \phi = g - \phi$ and $\partial_t \pi = g - \pi$.

We choose centered, second order accurate finite differences in space,
and the default time integrator is the classic fourth-order
Runge-Kutta method.  The numerical domain is an equidistant grid of
points in $[-\frac{L}{2},\frac{L}{2}]^d$, dimension $d=1$, $2$, or
$3$, with Cartesian coordinates. There are $N$ points in each of up to
three directions with a total of $V=N^d$ points.

Let us discuss some results for vanishing source term, $\rho = 0$, and
vanishing Dirichlet boundary, $g=0$, where the method has to reduce an
initial guess of $\phi = 1 / (1+x_j x^j)$ and $\pi = 0$ at $t=0$ to
the asymptotic, late-time value $\phi = \pi = 0$.

\begin{figure}[tb]
    \includegraphics[width=0.4\textwidth]{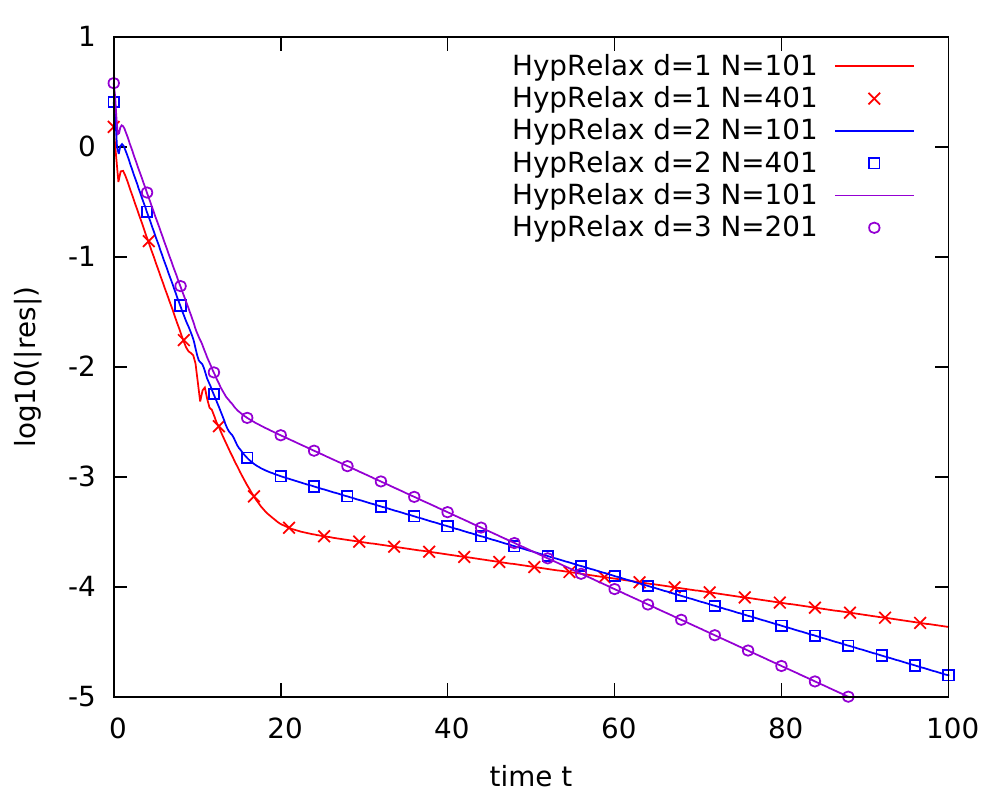}    
    \caption{
      \label{fig:PoissonFD1}
      Poisson equation, FD method.  Convergence of the residual with
      time for one, two, and three dimensions.  Shown is a solid line
      for $N=101$ and markers for a higher resolution given by $N=201$
      or $401$ points. On the scale of the plot, the convergence rate
      is independent of resolution for any given number of dimensions.
    }
\end{figure}

In Fig.~\ref{fig:PoissonFD1}, we show results for a box of size
$L=20$, damping parameter $\eta=0$, varying the number of points and
the number of dimensions. The norm is weighted by the grid spacing
$\Delta x$ to represent the integral of the residual, $|f|_2 = (\sum
f^2 \Delta x^d)^{1/2}$. Convergence is exponential in time, with two
distinct phases. Inspection of the evolution of $\phi$ and $\pi$ shows
that the initial phase corresponds to the damping of short wavelengths
(in this example until $t\approx 20$), after which long wavelengths
dominate and the convergence is slower.  The convergence of the
(weighted) norm of the residual with time is quite independent of the
resolution.  In this example the time-step is $\Delta t = \lambda
\Delta x$ for a fixed Courant factor $\lambda$, so the number of time
steps is proportional to the number $N$ of grid points in one
direction. The work per right-hand-side evaluation is $O(N^d)$, so the
total work to reach a final time $T$ is $O(N^{d+1})$.

\begin{figure}[tb]
    \includegraphics[width=0.45\textwidth]{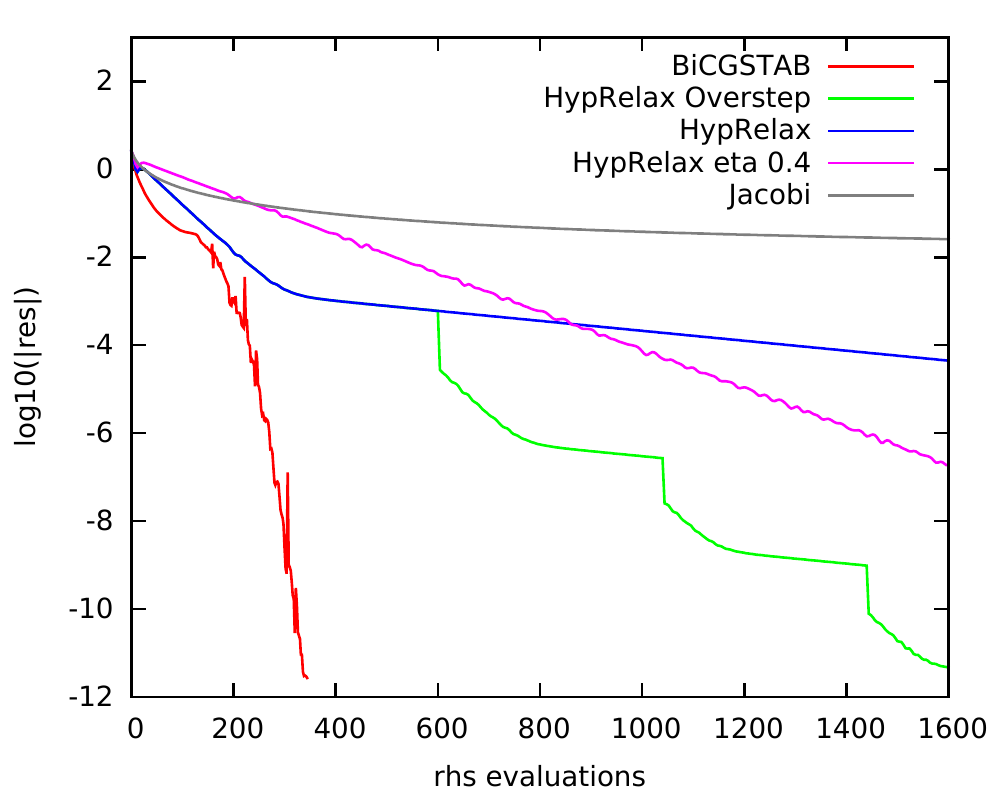}    
    \caption{
      \label{fig:PoissonFD2}
      Poisson equation, FD method. 
      Convergence of the residual with the number of
      right-hand-side evaluations. 
      Shown is a comparison between different methods for $N=101$ in two 
      dimensions.
    }
\end{figure}

A key question is how efficient hyperbolic relaxation is compared to
other methods. In Fig.~\ref{fig:PoissonFD2}, we show a comparison of
different methods for a two-dimensional example with $N = 101$ points.
The methods considered are hyperbolic relaxation as above, the
standard Jacobi iteration \cite{PreFlaTeu07a}, and the BiCGSTAB method
as an example for a Krylov subspace method \cite{BarBerCha93}. Also
included are two additional variants of hyperbolic relaxation. In
these examples $\Delta t = 1.0 \Delta x$ for RK4 in 2d.

Referring to Fig.~\ref{fig:PoissonFD2}, the Jacobi method shows the
slowest convergence. Reducing the residuum of the 2d Poisson equation
by a factor $10^{-p}$ requires $n \approx \frac{1}{2} p N^2$
iterations on a $N\times N$ grid \cite{PreFlaTeu07a}. For a 2d grid
with $V=N^2$ degrees of freedom, the operation count is therefore
$O(V^2)=O(N^4)$, compared to $O(N^3)$ for optimal SOR and $O(V\log V)$
for multigrid methods. Hyperbolic relaxation with $O(V)\times
O(N)=O(N^3)$, as demonstrated in Fig.~\ref{fig:PoissonFD1}, is
therefore a reasonable candidate for further consideration.  In the
concrete example, the Jacobi method is significantly slower than
hyperbolic relaxation, but the Jacobi method is usually not considered
as a stand-alone method.

For this simple comparison, the BiCGSTAB method is used without a
preconditioner, but the Laplace operator leads to a sufficiently well
conditioned operator such that convergence is fast nevertheless,
compared to the other methods considered here. There is an initial
phase of relatively slow convergence, but once the trial solution is
sufficiently close to the final answer, convergence becomes much
faster.

Remarkably, hyperbolic relaxation does about as well as BiCGSTAB
during the first phase. However, convergence slows down after the
shorter wavelengths have been damped and errors due to larger
wavelength remain. We have considered three ideas to improve the
convergence of hyperbolic relaxation for long wavelengths.  Not shown
here is the multi-level refinement strategy which we employ in the
\texttt{bamps} code, see Sec.~\ref{Subsection:RefinementStrategy}.

As an immediate application of the mode analysis of
Sec.~\ref{Subsection:ModeAna}, we introduced the damping parameter
$\eta$, which for the basic experiments so far was set to
$\eta=1$. Also shown in Fig.~\ref{fig:PoissonFD2} is the result for
$\eta=0.4$, which is slower than $\eta=1$ initially, but faster for
later iterations. This effect is related to the size $L$ of the
box. It seems possible to construct a dynamically adjusted damping
$\eta(t)$. Similar results hold for the parameter $b$ in
\eqref{hyprelax_b}. The velocity associated with the largest wave
number $k$ scales with $\sqrt{b}$ but is independent of $\eta$, so for
optimal performance the Courant factor has to be adjusted for the
version with $b$ but can be kept constant for the version with $\eta$.

We also experimented with a ``one-step overrelaxation'' method (as
opposed to successive overrelaxation).  This is based on the
observation that after the initial propagation/damping phase of
hyperbolic relaxation, the second time derivative of $\phi$ becomes
significantly smaller than the first time derivative,
$\partial_t^2\phi \ll \partial_t\phi$. Hence it seems promising to
attempt a {\em linear} extrapolation in time. The curve labeled
``overstep'' in Fig.~\ref{fig:PoissonFD2} is obtained by searching
every few iterations for the time step~$\Delta T = \lambda \Delta t$
that minimizes the global residual of~$\phi_{new} = \phi + \Delta T \,
F(\phi)$, where~$F$ is the update suggested by the time stepping
algorithm (\eg RK4).  This is similar to various other 1d step-size
optimizations. For the example considered here (but also for
$\rho\neq0$ as below), the late time solution of hyperbolic relaxation
is sufficiently regular that indeed an appropriate global $\Delta T$
can be found. The overstep algorithm only accepts improvements by a
given factor, say 10 (we tried 2 to 1000). After the adjustment the
solution is disturbed but converges again with the typical speed for
shorter wavelengths to a new regular state, so in the optimal case the
overall convergence rate approaches that of the fast phase of
hyperbolic relaxation.

The main points regarding the convergence rate of hyperbolic
relaxation as shown in Fig.~\ref{fig:PoissonFD2} are that the method
works out-of-the-box and that its performance falls somewhere between
Jacobi and BiCGSTAB. There seems to be quite some potential for
accelerating the convergence rate of hyperbolic relaxation. From the
point of view of solving elliptic equations with a code designed for
hyperbolic equations, note that hyperbolic relaxation is ``only''
slower by a factor of about 5 (to reach a residual of $10^{-9}$ in
this example) than a standard method like BiCGSTAB, which however may
not be readily available.

\subsection{Poisson Equation -- Pseudospectral Method}
\label{Subsection:Poisson}

To test the hyperbolic relaxation elliptic solver we start by solving
Poisson's equation,
\begin{equation}
\label{eq:Poisson}
\Delta \psi - \rho = 0\, ,
\end{equation}
in spherical symmetry, \ie $\rho = \rho(r)$, $r = \sqrt{x^i x_i}$. To
solve this equation we choose the hyperbolic relaxation system
\begin{align}
  \label{eq:PoissonHypRelaxPsi}
  \partial_t \psi &= \delta^{ij} \partial_i r_j - \rho \, ,\\
  \label{eq:PoissonHypRelaxR}
  \partial_t r_i  &= \partial_i \psi - r_i \, .
\end{align}
For our first test we take $\rho$ to be smooth, \ie it is infinitely
often continuous differentiable,
\begin{equation}
\label{eq:PoissonRhoGaussian}
\rho = \rho_0 \left ( \frac{-6}{R^2} + \frac{4 r^2}{R^4} \right )
e^{-r^2/R^2} \, ,
\end{equation}
where $R$ and $\rho_0$ are non-zero parameters. For this $\rho$
Poisson's equation has the solution
\begin{equation}
\label{eq:PoissonSolutionGaussian}
\psi_{\rm analytic} = \rho_0 e^{-r^2/R^2} \, .
\end{equation}
At the boundary a falloff in $\psi$ compatible with this solution is
obtained by imposing the Robin boundary condition $\partial_r \psi =
s^i \partial_i \psi = {- 2 r}\psi/{R^2}$.

For our second test we take a non-smooth $\rho$ that corresponds to a
homogeneously charged sphere, which is like a toy model for stars.
The density $\rho$ is then given by
\begin{equation}
\label{eq:PoissonRhoSphere}
\rho = 
\begin{cases}
\rho_0 & \text{ if } r \leq R \\ 
0  & \text{ if } r > R \, ,
\end{cases}
\end{equation}
for which the Poisson equation has the solution
\begin{equation}
\label{eq:PoissonSolutionSphere}
\psi_{\rm analytic} = \rho_0
\begin{cases}
  \frac{r^2}{6} - \frac{R^2}{2} & \text{ if } r \leq R \\ 
  -\frac{R^3}{3 r} & \text{ if } r > R \, .
\end{cases}
\end{equation}
Again we impose Robin boundary conditions according to the falloff of
this solution, \ie $\partial_r \psi = s^i \partial_i \psi = -\psi/r$.

In our tests we place the outer boundary at radius of 10 and and we
divide the grid into a total of eight subpatches, where the inner five
extend over the interval $[0,5]$ and the outer three, having a coarser
resolution, extend over $[5,10]$.  The parameters determining $\rho$
are chosen to be $R = 5$ and $\rho_0 = 1$.  For the non-smooth case
special care has to be taken to ensure convergence.  In particular we
chose the grid such that the discontinuity lies at a boundary of
subpatch, ensuring second order convergence.  In both test cases the
relaxed solution converges with the number of grid points to the
analytical solution. To investigate the convergence we have to make
sure that the solution is completely relaxed on every resolution.
This is achieved by choosing in Eq.~\eqref{eq:RefineCriterion} $c =
0.0001$.  In Fig.~\ref{fig:PoissonConv} we report the absolute
difference between the analytical and numerical solution integrated
over the outermost subpatch. We note however that the convergence
behavior is the same on all other subpatches.  As expected we find the
error of the numerical solution to decrease exponentially with the
number of points for the smooth $\rho$ from
Eq.~\eqref{eq:PoissonRhoGaussian}. For the non-smooth $\rho$ of
Eq.~\eqref{eq:PoissonRhoSphere}, we only get a convergence order of
approximately two, which is the expected convergence order for
discontinuous $\rho$.  Of course this is not very efficient for a
spectral method. For non-smooth right-hand sides it might be
preferable to increase the number of subpatches (h-refinement) instead
of the number of collocation points per subpatch.

\begin{figure}[tb]
    \includegraphics[width=0.5\textwidth]{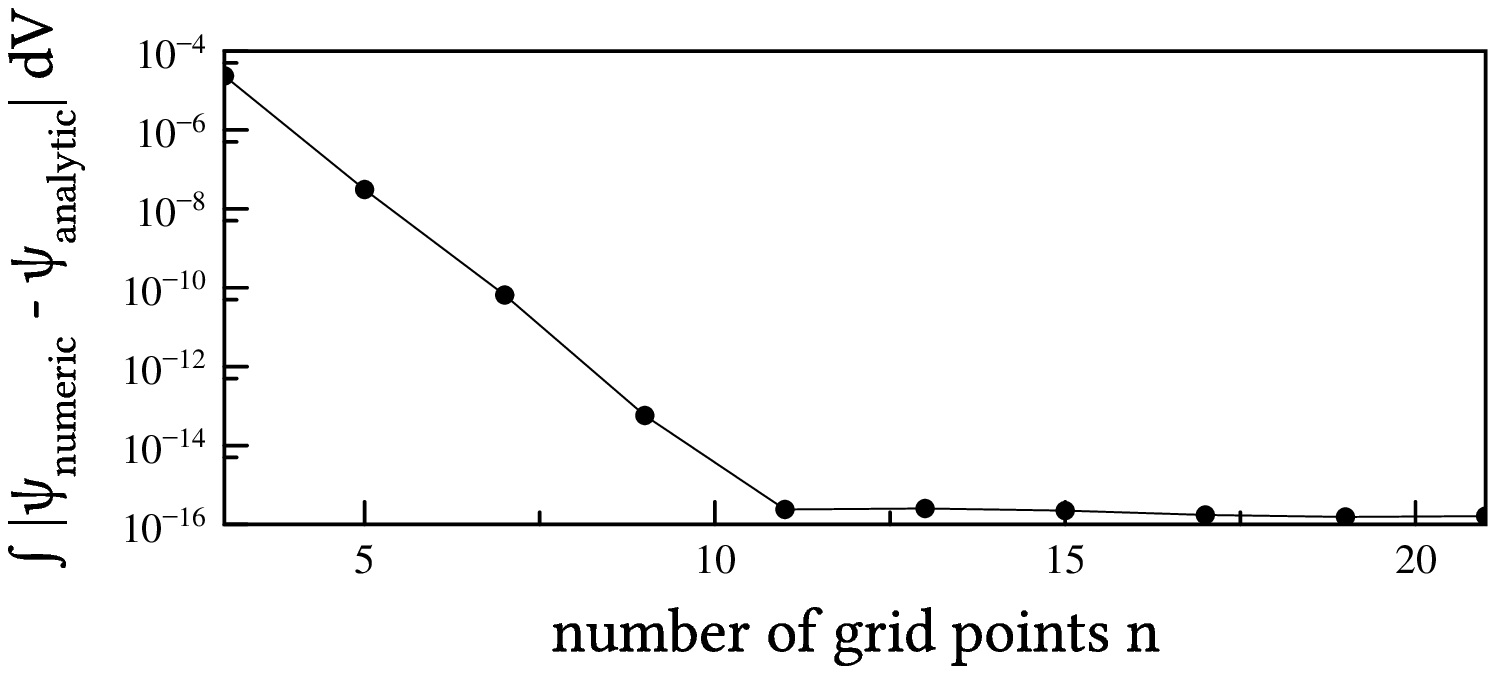}    
    \includegraphics[width=0.5\textwidth]{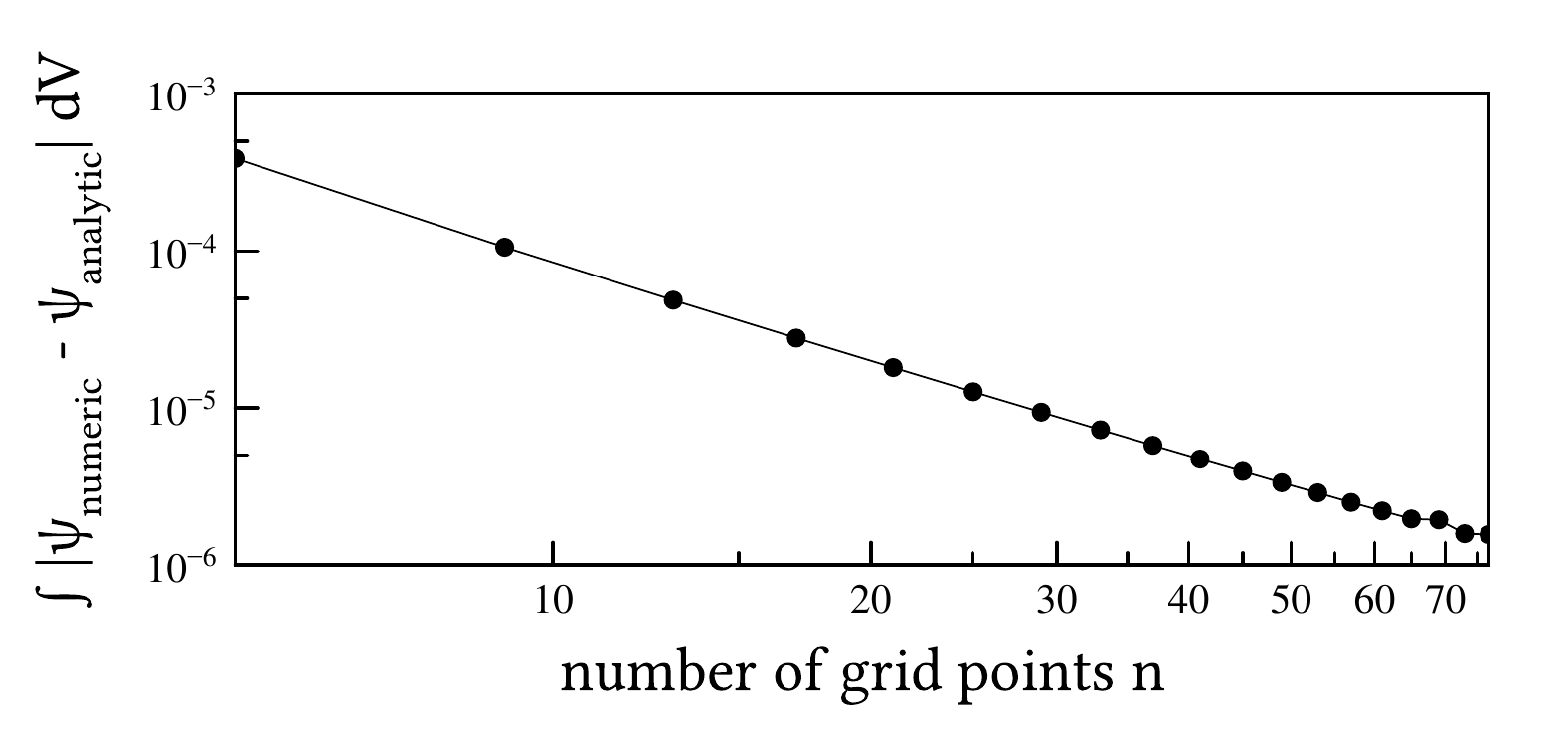}    
    \caption{
      \label{fig:PoissonConv} Convergence L1-norm of the difference 
      between the analytical and the numerical solution.  Upper plot:
      for smooth $\rho$ (Eq.~\eqref{eq:PoissonRhoGaussian}).  Lower
      plot: for non-smooth $\rho$ (Eq.~\eqref{eq:PoissonRhoSphere}).
      Note that in the upper plot only the error axis is logarithmic,
      while in the lower plot both axes are logarithmic.  }
\end{figure}

In Fig.~\ref{fig:PoissonRelaxation} we investigate how the L1-norm of
different quantities, that can be used to approximate the error,
progresses during the relaxation process.  First we observe that the
difference to the analytical solution decreases even when the computed
residual, given by left-hand side of Eq.~\ref{eq:Poisson}, is already
leveling off. This is especially remarkable for the non-smooth case,
where the residual itself is not converging at all.  For the smooth
case we secondly observe that after refining the grid the norm of
right-hand side of Eq.~\eqref{eq:PoissonHypRelaxPsi} practically
continues at the same level as before. The norm of the residual on the
other hand drops quickly after refining, reaching the right-hand sides
level until again the discretization limit is reached.  These
observations suggest that for problems with smooth solutions it is
preferable to relax for longer on the coarse grid.  For problems with
non-smooth solutions, however, more new error develops during each
refinement and thus refining for longer on the coarse grid is not
paying off. Furthermore, it is preferable to increase the grid
resolution faster.

\begin{figure}[tb]
    \includegraphics[width=0.5\textwidth]{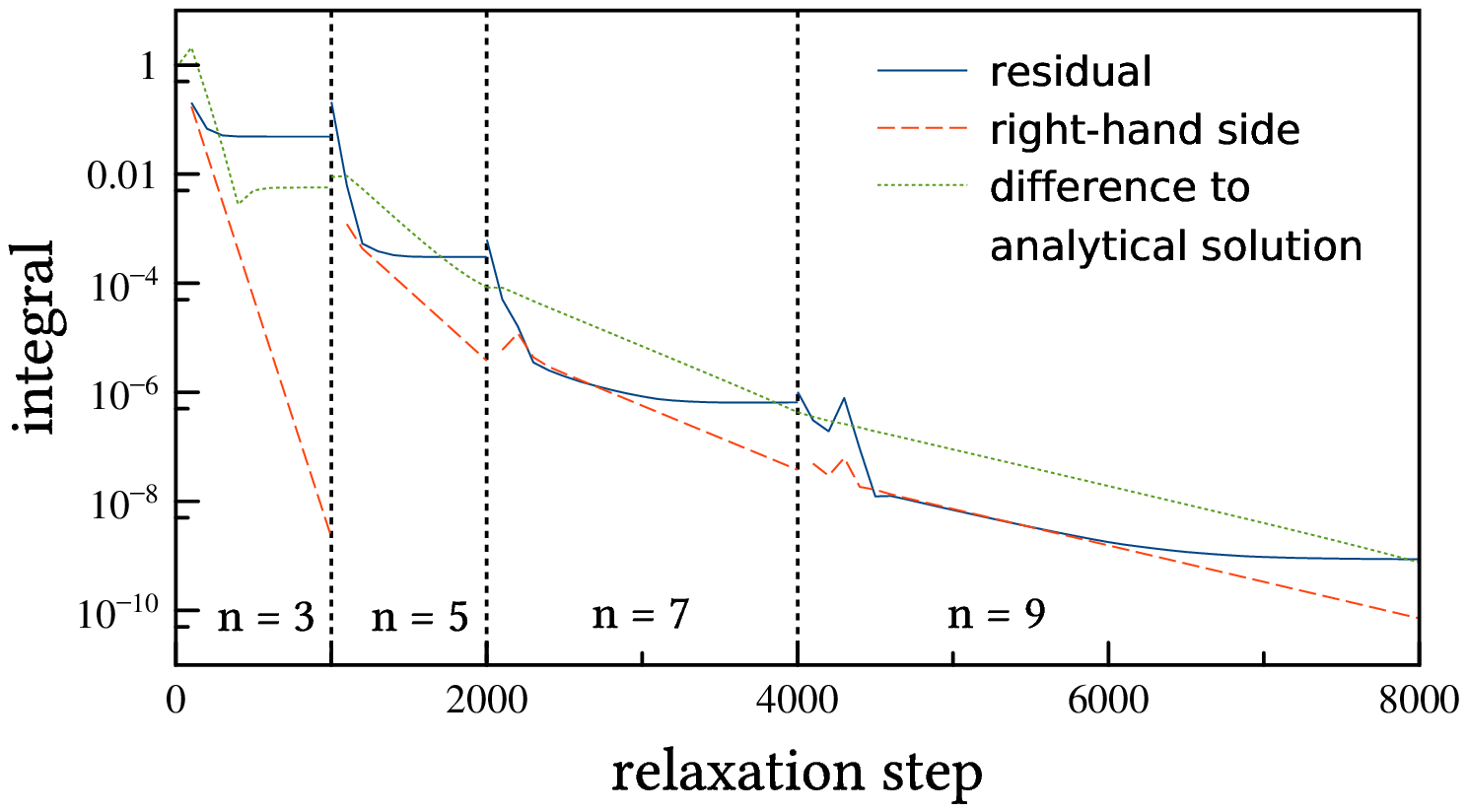}    
    \includegraphics[width=0.5\textwidth]{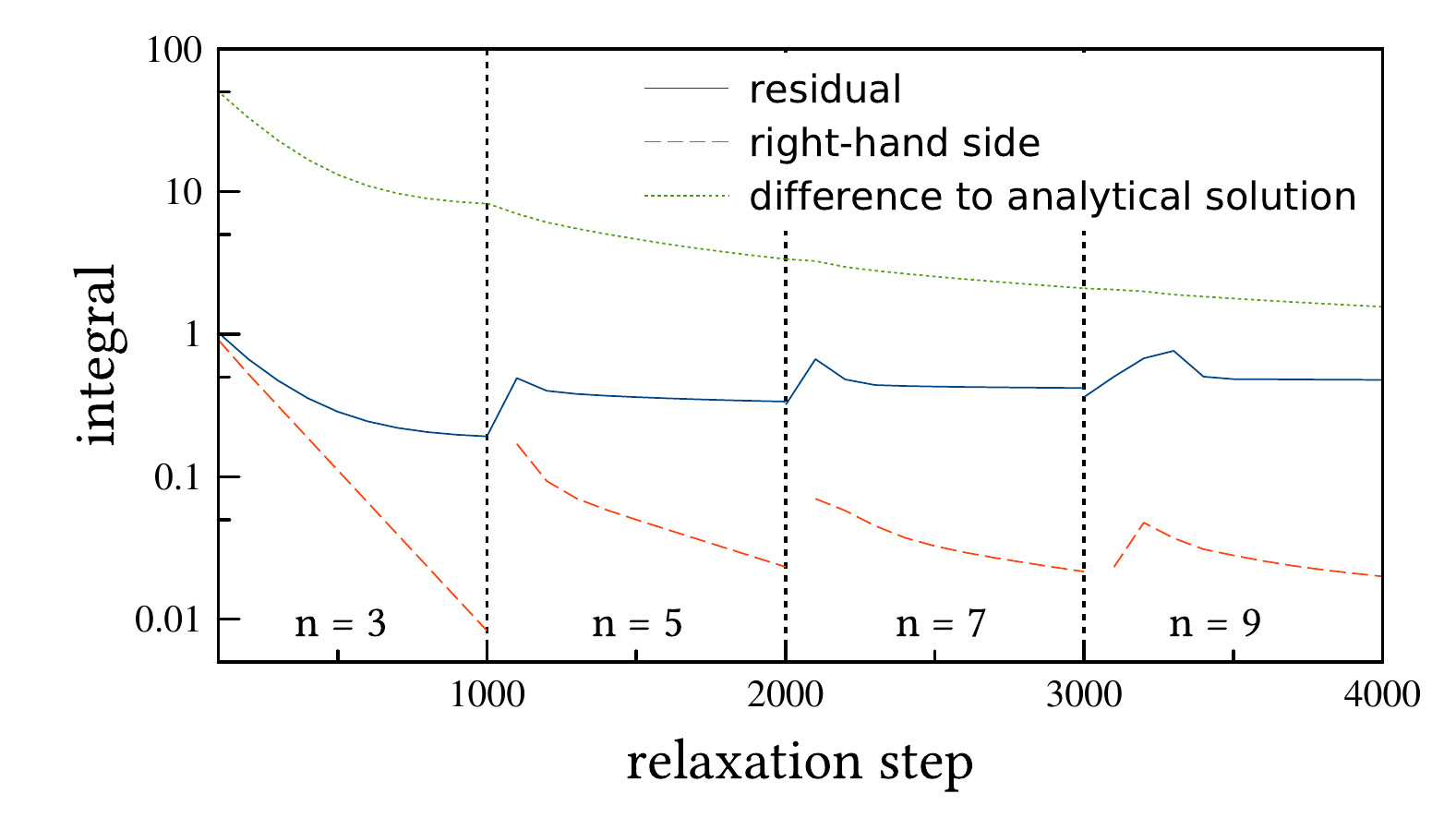}    
    \caption{
      \label{fig:PoissonRelaxation} Progression of the L1-norm of different error quantities
      during of the relaxation process for the Poisson
      equation. Vertical dashed lines indicate transitions to a finer
      grid.  The respective error quantities are: blue solid line:
      residual, defined as left-hand side of Eq.~\ref{eq:Poisson},
      orange dashed line: right-hand side of
      Eq.~\ref{eq:PoissonHypRelaxPsi}, green dotted line: difference
      to the analytical solution.  Upper plot: for smooth $\rho$
      (Eq.~\eqref{eq:PoissonRhoGaussian}).  Lower plot: for non-smooth
      $\rho$ (Eq.~\eqref{eq:PoissonRhoSphere}).  }
\end{figure}

As a last simple test, we investigated the behavior in the case of
non-unique solutions. For this we took the smooth $\rho$ from
Eq.~\eqref{eq:PoissonRhoGaussian} and imposed the Neumann boundary
condition $\partial_r \psi = 0$, for which multiple solutions
differing only by an additive constant exist. We find that after some
relaxation the right hand side of Eq.~\eqref{eq:PoissonHypRelaxPsi}
becomes approximately constant in space. From this point on the
solution is no longer improving, since only constant terms, which do
not improve the residual of Eq.~\eqref{eq:Poisson}, are added.

\section{Application to Initial Data for General Relativity}
\label{Section:Applications}
\subsection{The Extended Conformal Thin-Sandwich Equations}
\label{Subsection:XCTS}

In numerical relativity one usually decomposes the spacetime metric
$g_{ab}$ into a temporal and spatial part in the form
\begin{equation}
\label{eq:PhysMetric}
g_{ab} d x^a dx^b = -\alpha^2 d t^2 + \gamma_{ij} (d x^i + \beta^i d
t) (d x^j + \beta^j d t) \, ,
\end{equation}
where $\alpha$ is called the lapse, $\beta^i$ the shift and
$\gamma_{ij}$ the spatial metric.  The equations of motion in general
relativity are subject to constraint equations, which have to be
solved before the spacetime is evolved numerically.  A popular
formulation of the constraint equations is given by the extended
conformal thin-sandwich (XCTS) equations~\cite{BauSha10,Tic17} in
which the spatial metric is decomposed into a conformal factor $\psi$
and a spatial conformal metric $\bar \gamma_{ij}$ as $\gamma_{ij} =
\psi^4 \bar \gamma_{ij}$.  In the XCTS framework the constraint
equations take the form
\begin{align}
 \label{eq:XCTSpsi}    
  \bar D^j \bar D_j \psi =& \frac{\psi}{8} \bar R - \psi^5 \left ( 2
  \pi \rho - \frac{K^2}{12} + \frac{1}{8} A_{ij} A^{ij} \right ) \,,\\
  \begin{split}
  \label{eq:XCTSbeta}
  \bar D^j \bar D_j \beta^i =& - \frac{1}{3} \bar D^i \bar D_j \beta^j
  - \bar R\indices{^i_j} \beta^j + 16 \pi \alpha \psi^4 J^i \\
  & + (
  \bar D^i \beta^j + \bar D^j \beta^i - \frac{2}{3} \bar \gamma^{ij}
  \bar D_k \beta^k ) \\
  & \quad \cdot \bar D_j \ln(\alpha \psi^{-6})
  \\
  & - \alpha \psi^{-6} \bar D_j ( \alpha^{-1} \psi^6 \partial_t
  \bar \gamma^{ij} ) + \frac{4}{3} \alpha \bar D^i K \, ,
 \end{split} \\
 \begin{split}
 \label{eq:XCTSpsialpha}  
  \bar D^j \bar D_j ( \alpha \psi ) =& \alpha \psi^5 \left
  (\frac{7}{8} A_{ij} A^{ij} + \frac{5}{12} K^2 + 2 \pi (\rho + 2 J)
  \right ) \\
  & - \psi^5 (\partial_t - \beta^j \bar D_j ) K +
  \frac{1}{8} \alpha \psi \bar R \, .
 \end{split}
\end{align}
Here $\bar D_i$ is the covariant derivative compatible with the
conformal metric $\bar \gamma_{ij}$, $\bar R\indices{_i_j}$ is the
Ricci tensor of $\bar \gamma_{ij}$ and $\bar R$ is the corresponding
Ricci scalar.  The tensor $A_{ij}$ is the tracefree part of the
extrinsic curvature $K_{ij}$ and $K$ is the trace of $K_{ij}$.  The
matter source terms are defined as the following contractions of the
energy-momentum tensor $T_{ab}$: $\rho = T_{ab} n^a n^b$, $J^i =
-T_{ab} \gamma^{i a} n^b$ and $J = \gamma^{ab} T_{ab}$, where $n^a$ is
the timelike vector normal to the spatial hypersurface $n^a =
(1/\alpha,-\beta^i/\alpha)$. For conformal quantities (denoted by a
bar) the conformal spatial metric $\bar \gamma^{ij}$ lowers and raises
indices and for unbarred quantities the physical spatial metric
$\gamma^{ij}$ is used.  In the XCTS equations $\bar \gamma_{ij}$,
$\partial_t \bar \gamma_{ij}$, $K$ and $\partial_t K$ are given
functions, depending on the type of initial data you want to
construct.

In Eq.~\eqref{eq:XCTSpsialpha} the product $\alpha \psi$ is taken as
one variable. For our computations we rewrite this equation with the
help of Eq.~\eqref{eq:XCTSpsi} as
\begin{equation}
 \begin{split}
 \label{eq:XCTSalpha}  
  \bar D^j \bar D_j \alpha =& -\frac{2}{\psi} ( \bar D^j \alpha) (\bar
  D_j \psi) - \psi^4 (\partial_t - \beta^j \bar D_j ) K \\
  & + \alpha
  \psi^4 \left (A_{ij} A^{ij} + \frac{K^2}{3} + 4 \pi (\rho + J)
  \right ) \, ,
 \end{split}
\end{equation}
so we can solve directly for $\alpha$.  We have tested our elliptic
solver with both versions and found them to work equally well in our
applications. In the following we investigate the XCTS system with our
replacement for the lapse equation, but the analysis would be exactly
the same for the original system.

The principal part of the XCTS equations is given by
\begin{equation}
 \label{XCTSPrincPart}
  \begin{pmatrix}
  \bar \gamma^{kj}  &  0  &  0  \\
  0  &  \bar \gamma^{kj} \delta^l_i + \frac{1}{3} \bar \gamma^{l k} \delta^j_i   & 0  \\
  0  &  0  &  \bar \gamma^{kj}  \\
  \end{pmatrix} 
  \partial_k \partial_j 
  \begin{pmatrix}
  \psi \\
  \beta^i \\
  \alpha \\
  \end{pmatrix} 
  \, .
\end{equation}
The principal part is coupled only between the components of
$\beta^i$. Thus we can carry out the hyperbolicity analysis
independently for $\psi$, $\alpha$ and $\beta^i$.

The metric to lower and raise spatial indices in the hyperbolic
relaxation method is in principal arbitrary, but for the XCTS
equations we use the conformal metric $\bar \gamma_{ij}$, because this
choice simplifies the following formulas considerably. Another
peculiarity is the fact that spatial indices and field indices
``mix'', but in general they are lowered and raised by different
metrics. Therefore we do not lower and raise field indices and instead
write the Euclidian metric explicitly.

For the conformal factor part we have in the hyperbolic relaxation
system $a^{ij} = \bar \gamma^{ij}$ and we choose $b\indices{^i_j} =
\delta^i_j$. Here we have suppressed the field indices, because we
only have a single field $\psi$.  The characteristic variables and
speeds of the hyperbolic relaxation system of the $\psi$ part are then
given by
\begin{align}
 \label{eq:XCTSPsiCharacteristics0}
\hat u^0_j &=  r_j - s_j s^i r_i^{(\psi)} \, , 
~~~~~~~~~~~  \lambda^0_{i} = 0 \, , \\
\label{eq:XCTSPsiCharacteristics+-}
\hat u^{\pm} &= \psi \pm s^i r_i^{(\psi)} \, ,
~~~~~~~~~~~~~\, \lambda^\pm = \pm 1 \, ,
\end{align}
where $r_i^{(\psi)}$ is the reduction variable for $\psi$.

The lapse part has exactly the same principal part, so we have
identical $a^{ij}$ and $b\indices{^i_j}$ and the characteristics are
\begin{align}
 \label{eq:XCTSAlphaCharacteristics0}
\hat u^0_j &=  r_j - s_j  s^i r_i^{(\alpha)} \, ,
~~~~~~~~~~~  \lambda^0_{i} = 0 \, , \\
\label{eq:XCTSAlphaCharacteristics+-}
\hat u^{\pm} &= \alpha \pm \bar s^i r_i^{(\alpha)} \, ,
~~~~~~~~~~~~~\, \lambda^\pm = \pm 1 \, ,
\end{align}
with $r_i^{(\alpha)}$ being the reduction variable for $\alpha$.

In the derivation of the hyperbolic relaxation equations we labeled
the fields with lower indices. For the shift however we use upper
indices here and thus one has to be careful, not to confuse the
indices. Therefore, we introduce auxiliary fields $\phi_\alpha$ with
\begin{equation}
 \beta^\alpha = \delta^{\alpha \beta} \phi_\beta \, .
\end{equation}
Substituting $\beta^\alpha$ we obtain for the principal part of the 
shift equations
\begin{equation}
 a\indices{^i^j^\beta_\alpha} \partial_i \partial_j \phi_\beta = (
 \bar \gamma^{i j} \delta^\beta_\alpha + \frac{1}{3} \bar \gamma^{i
   \epsilon} \delta_{\epsilon \alpha} \delta^{j \beta} ) \partial_i
 \partial_j \phi_\beta \, .
\end{equation}
We take $b\indices{^i_j^\beta_\alpha}$ to be the inverse of $a$ (as
defined in section~\ref{Subsection:HyperbolicityAnalysis}),
$b\indices{^i_j^\beta_\alpha} = \delta^i_j \delta^\beta_\alpha -
\frac{1}{6} \delta^{i \beta} \delta_{j \alpha} $ .  The characteristic
variables and speeds for this part of the hyperbolic relaxation system
are then given by
\begin{align}
\label{eq:XCTSBetaCharacteristics0}
\hat u^0_{j\gamma} &= r_{j\gamma} - b\indices{^s_j^\epsilon_\gamma} (
     {b\indices{^s_s}}^{-1} )\indices{^\alpha_\epsilon} s^i
     r_{i\alpha} \\ &= r_{j\gamma} - s_j s^i r_{i \gamma} +
     \frac{1}{5} (\delta_{j \gamma} - s_j s^l \delta_{l \gamma}) s_k \delta^{k \alpha} s^i r_{i
       \alpha} \, ,\\
     \lambda^0_{i\gamma} &= 0 \,, \\
\label{eq:XCTSBetaCharacteristics+-}
\hat u^{\pm}_\gamma &= \phi_\gamma \pm a\indices{^s^i^\alpha_\gamma}
r_{i\alpha} ~~~~~~ \\
&= \phi_\gamma \pm (s^i r_{i\gamma} +
\frac{1}{3} s^\epsilon \delta_{\epsilon \gamma} \delta^{j \alpha} r_{j
  \alpha}) \, , \\
\lambda_\gamma^\pm &= \pm 1 \, ,
\end{align}
where the $r\indices{_i_\alpha}$ denote the reduction variables for
the auxiliary fields $\phi_\alpha$

At the domain boundary we want the solution to fall off like the
Schwarzschild solution, \ie $\psi = \frac{a}{r} + 1$ and $\alpha =
\frac{b}{r} + 1$.  This ansatz gives rise to the following Robin
boundary conditions
\begin{equation}
 \label{eq:XCTSBCpsialpha}
 \partial_s \psi |_{\partial \Omega} = \frac{1-\psi}{r} \, , ~~~~~~
 \partial_s \alpha |_{\partial \Omega} = \frac{1-\alpha}{r} \, .
\end{equation}
For the shift we likewise impose a radial falloff by the Robin
condition
\begin{equation}
 \label{eq:XCTSBCbeta}
 \partial_s \beta^i |_{\partial \Omega} = \frac{\beta^i}{r} \, .
\end{equation}

As an initial guess we always use the flat space solution, \ie $\psi =
1$, $\alpha=1$, $\beta^i = 0$.  Of course an initial guess, that is a
good approximation to the solution is always the preferred start for
the relaxation, since it will take less time to relax to the solution
or might be necessary to relax at all. However we find our simple
initial guess to work well and it demonstrates in a nice way the high
robustness of the hyperbolic relaxation method exhibited in our
experiments.

\subsection{Scalar Field}\label{Subsection:ScalarField}

The energy-momentum tensor for a scalar field $\phi$ is given by 
\begin{equation}
 \label{eq:ScalarFieldEMTensor}
 T_{ab} = \nabla_a \phi \nabla_b \phi - \frac{1}{2} g_{ab} \left (
 \nabla_c \phi \nabla^c \phi + m^2 \phi^2 \right ) \, ,
\end{equation}
where $\nabla$ denotes the covariant derivative compatible with
$g_{ab}$.  We consider conformally flat moment-of-time-symmetry
initial data, \ie $\bar \gamma_{ij} = \delta_{ij}$, $n^a \nabla_a \phi
= 0$, $\partial_t \bar \gamma_{ij} = 0$ and maximal slicing $K=0$,
$\partial_t K = 0$. This yields for the matter quantities
\begin{align}
 \label{eq:ScalarFieldRho}
 \rho &= \frac{1}{2} \gamma^{ij} (\partial_i \phi) (\partial_j \phi) +
 \frac{1}{2} m^2 \phi^2 \, , \\
  \label{eq:ScalarFieldJi}
 J^i &= 0 \, , \\
  \label{eq:ScalarFieldJ}
 J &= -\frac{1}{2} \gamma^{ij} (\partial_i \phi) (\partial_j \phi) -
 \frac{3}{2} m^2 \phi^2 \, .
\end{align}
For a massless scalar field ($m=0$) the solutions for lapse and shift
are trivially given by $\alpha = 1$ and $\beta^i = 0$ and we only have
to determine the conformal factor $\psi$ by solving
Eq.~\eqref{eq:XCTSpsi}, now taking the form
\begin{equation}
 \label{eq:ScalarFieldXCTSpsi}
 0 = \delta^{ij} \partial_i \partial_j \psi + \pi \psi \delta^{ij}
 (\partial_i \phi) (\partial_j \phi) \, .
\end{equation}
For a massive scalar field ($m \neq 0$) one would have either have to
solve additionally for the lapse or one gives up the requirement on
$\partial_t K$.  For the scalar field we choose radially symmetric,
smooth initial data of the form
\begin{equation}
 \label{eq:ScalarFieldphi}
 \phi(r) = p \left(\tanh \frac{r}{\sigma } - \tanh \frac{r}{\sigma } \right)  \, ,
\end{equation}
where $\sigma$ and $p$ are free parameters, that we choose for our
test to be $\sigma = 1$ and $p=0.1$.

The computational domain is divided into eight subpatches, where the
inner five subpatches are of smaller extent to improve the resolution
near the center. In Fig.~\ref{fig:ScalarFieldConfFactor} we show the
numerical solution for the conformal factor and the absolute value of
the residual of Eq.~\eqref{eq:XCTSpsi} for a resolution of 21
collocation points per subpatch.  A general feature that can be
observed in the residual are spikes at the boundary of subpatches,
which are expected for quantities involving first and second
derivatives in a discontinuous Galerkin approximation.

We also investigate how the solution converges with increasing
resolution. For this purpose we look at the Chebyshev coefficients and
investigate their convergence against the coefficients of a high
resolution solution. In Fig.~\ref{fig:ScalarFieldConv} we present the
convergence behavior of the lowest Chebyshev mode $C(0,0,0)$ against
its value for a resolution of 31 collocation points. We observe an
exponential convergence until we hit machine precision at around a
resolution of 25 collocation points per subpatch.

\begin{figure}[tb]
    \includegraphics[width=0.5\textwidth]{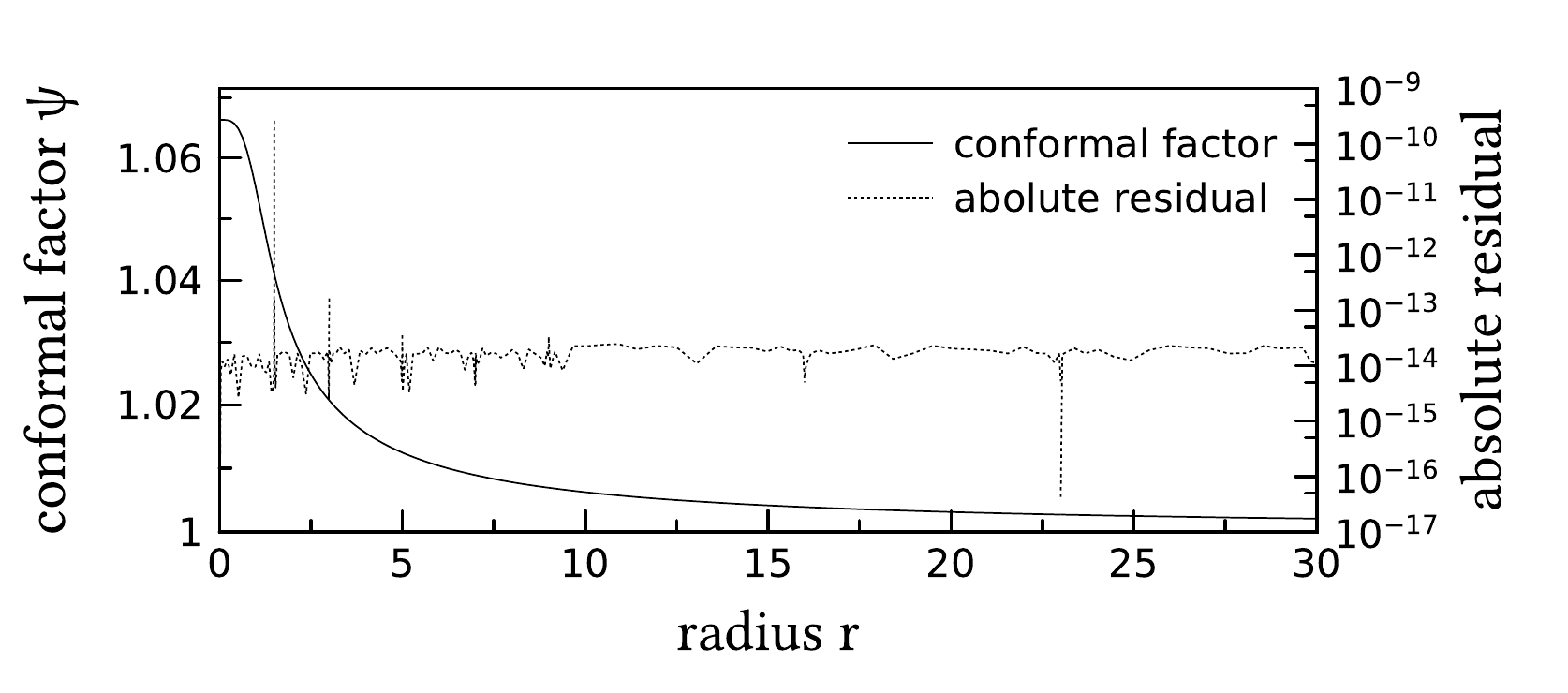}   
    \caption{
      \label{fig:ScalarFieldConfFactor} 
      Steady state at a resolution of 21 collocation points per
      subpatch for the initial data of the scalar field.  Solid line:
      conformal factor. Dashed line: residual for conformal factor, as
      given by the right-hand side of
      Eq.~\eqref{eq:ScalarFieldXCTSpsi}.  }
\end{figure}
\begin{figure}[tb]
    \includegraphics[width=0.5\textwidth]{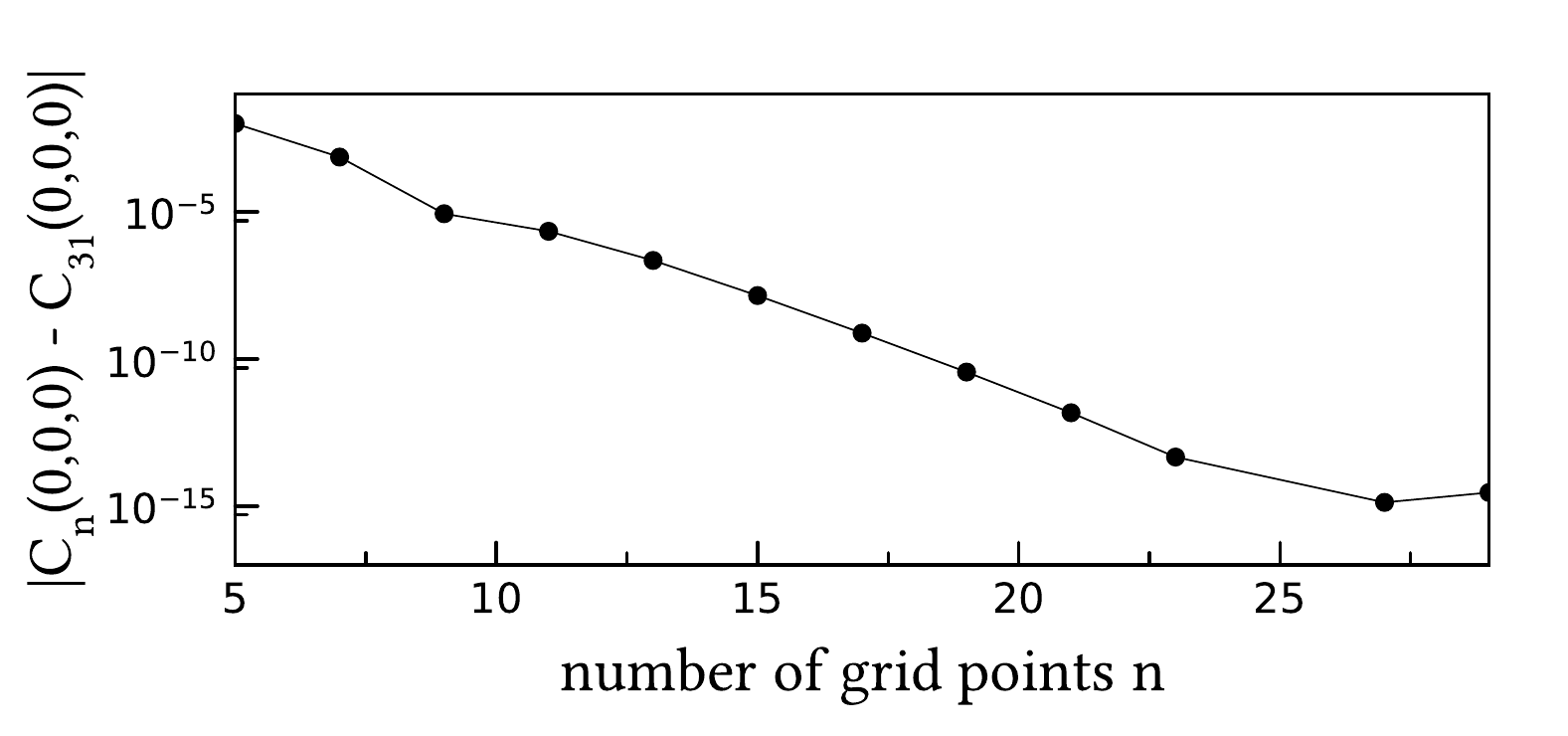}   
    \caption{
      \label{fig:ScalarFieldConv} 
      Convergence of the lowest Chebyshev coefficient $C(0,0,0)$ at
      the innermost subpatch for the initial data of the scalar field.
      The system is relaxed for different numbers of collocation
      points.  The plot shows the absolute value of the difference
      between the Chebyshev coefficient at the highest resolution (31
      collocation points) and its value for $n$ collocation points per
      subpatch.  For $n = 25$ the difference is zero within machine
      precision and is therefore not displayed on the logarithmic
      axis.  }
\end{figure}
\subsection{Tolman-Oppenheimer-Volkoff Star}\label{Subsection:TOVstar}

The energy-momentum tensor of a perfect fluid is given by
\begin{equation}
 T_{ab} = (\epsilon + p) u_a u_b + p g_{ab} \, ,
\end{equation}
where $\epsilon$ is the proper energy density, $p$ the fluid pressure
and $u_a$ the fluid four-velocity.  The Tolman-Oppenheimer-Volkoff
(TOV) solution~\cite{Mol12, MolMarJoh14, BauSha10} is a static
radially symmetric solution to general relativity, thus we have $u_a =
(\alpha, 0)$, $\bar \gamma_{ij} = \delta_{ij}$, $\partial_t \bar
\gamma_{ij} = 0$ and we consider again maximal slicing $K=0$,
$\partial_t K = 0$.  The matter quantities then become
\begin{equation}
 \label{eq:TOVmatter}
 \rho = \epsilon \, , ~~~~~~ J^i = 0 \, , ~~~~~~ J = 3 p   \, .
\end{equation}
We assume for our tests a polytropic equation of state and express the
matter quantities in terms of the specific enthalpy $h$,
\begin{align}
 \rho &= \left ( 1 + \frac{n(h-1)}{1+n} \right ) \left (
 \frac{h-1}{\kappa (1+n)} \right )^n \, , \\
 J &= 3 \kappa \left (
 \frac{h-1}{\kappa (1+n)} \right )^{n+1} \, ,
\end{align}
where $\kappa$ is the adiabatic constant and $n$ is the polytropic
index.  The Euler equation follows from energy-momentum conservation,
\ie for a fluid with temperature $T=0$
\begin{equation}
 u^a ( \nabla_a (h u_b) - \nabla_b (h u_a) ) = 0 \, ,
\end{equation}
which has to be satisfied in addition to the XCTS equations. For our
assumptions the Euler equation is satisfied for $\alpha h = const.$
and thus specifying the specific enthalpy at the origin yields
\begin{equation}
 h(r) = 
 \begin{cases}
    \frac{h(0) \alpha(0)}{\alpha(r)} & \text{ if } \alpha(r) < h(0) \alpha(0) \\ 
    1 & \text{ else } 
 \end{cases} \, .
\end{equation}
We can immediately get the solution for the shift $\beta^i = 0$ and
are left with solving Eq.~\eqref{eq:XCTSpsi} and~\eqref{eq:XCTSalpha}
of the XCTS system.

For our test we choose an adiabatic constant of $\kappa = 123.6489$
and a polytropic index of $n = 1$ and the enthalpy in the star's
center is set to $h(0) = 1.2$.  We present the solution for the
conformal factor and the lapse in Fig.~\ref{fig:TOVstarPsiAlpha} and
investigate the convergence of their lowest Chebyshev modes in
Fig.~\ref{fig:TOVstarConv}.  The residuals are large at the stellar
surface, \ie where $\alpha(r) = h(0) \alpha(0)$.  This is caused by
the fact that the matter terms are not smooth at this point, which can
be seen in the kink in the specific enthalpy $h$.  As for the
non-smooth right-hand sides discussed in
Sec.~\ref{Subsection:Poisson}, we have to make sure that the kink lies
at a subpatch boundary. By trial and error we find the stellar surface
for the above parameters to be located at $r = 9.7098$ and we place a
subpatch boundary at this position ``by hand''.  A more sophisticated
method is to fit coordinates automatically to the stellar
surface~\cite{Tic09a}. This however is beyond the scope of this paper
as we here want to focus on applications of the hyperbolic relaxation
method.  In contrast to what we observed in the non-smooth case for
the Poisson equation, here the Chebyshev coefficients converge
exponentially despite the kink in the specific enthalpy. The
convergence rate however is much smaller than that observed for the
initial data of the scalar field.

\begin{figure}[tb]
    \includegraphics[width=0.5\textwidth]{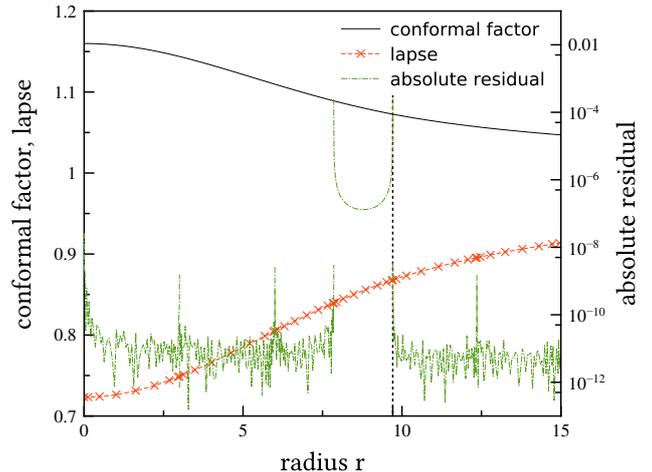}   
    \caption{
      \label{fig:TOVstarPsiAlpha} 
      Steady state at a resolution of 27 collocation points per
      subpatch for the initial data of the TOV star.  Solid line:
      conformal factor.  Orange dashed line with markers: lapse.
      Green dash-dotted line: absolute value of the residual for the
      conformal factor, as given by the right-hand side of
      Eq.~\eqref{eq:XCTSpsi}.  Vertical dashed line: position of the
      stellar surface.  }
\end{figure}
\begin{figure}[tb]
    \includegraphics[width=0.5\textwidth]{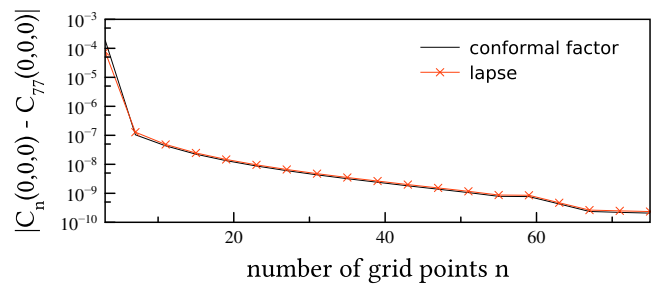}   
    \caption{
      \label{fig:TOVstarConv} 
      Convergence of the lowest Chebyshev coefficient as in
      Fig.~\ref{fig:ScalarFieldConv} for the initial data of the TOV
      star. The highest resolution used in these runs was 77
      collocation points per subpatch.  }
\end{figure}
\subsection{Neutron Star Binaries}\label{Subsection:NeutronStarBinaries}

For the construction of neutron star binary initial data we follow the
scheme of~\cite{MolMarJoh14} using the constant three-velocity
approximation. However, to solve the XCTS equations we do not rely on
iterating the solution of the equations for the conformal factor,
lapse and shift separately, but instead we solve for all variables
simultaneously relaxing the complete XCTS system, which accelerates
the solution process.  This is a feature that most solvers for this
type of initial data do not provide, and it could turn out to be an
advantage of our method.  Furthermore, we do not start with superposed
(boosted) TOV solutions, but instead start with a flat metric $\psi =
1$, $\alpha = 1$, $\beta^i$, as discussed at the end of
Sec.~\ref{Subsection:XCTS}.

For our test we consider equal mass neutron stars with a specific
enthalpy of $h = 1.01$ in each of their centers and a separation of
80. For the equation of state we choose again $\kappa = 123.6489$ and
$n = 1$. The stars' centers are located at the $z$-axis and their
velocities are parallel to the $x$-axis. We construct in initial data
for irrotational stars on a quasicircular orbit.  In
Fig.~\ref{fig:BNSPsiAlphaBeta} we present results for the conformal
factor, lapse, the $x$-component of the shift, the residual of the
conformal factor equation.  As for the TOV star initial data, the
residuals are biggest on those subpatches which contain the stellar
surfaces, where the matter fields are not smooth.

Because we are not using surface-fitted
coordinates~\cite{BonGouMar98a,Ans06,Tic09a} we can not place the
stellar surface at a subpatch boundary, and thus no high-order
convergence in the norm of the residuals can be seen with increasing
number of collocation points.  Although this may be a disadvantage for
studies of initial data per se, the situation changes if the goal is
evolution of the data. Since in an actual evolution of this data
surface-fitted coordinates are normally not retained, the high
accuracy of initial data with surface-fitted coordinates will be lost
relatively quickly anyway. On the other hand, methods like
\cite{Tic09a} require expensive iterations to determine the surface
fitting coordinates as part of the solution process, so any method
which works without special coordinates is more efficient in that part
of the algorithm. In fact, part of the motivation for the multigrid
method in \cite{MolMarJoh14} was to construct a solver which works on
a general Cartesian grid without surface-fitting coordinates. The
hyperbolic relaxation method achieves the same goal.

\begin{figure}[tb]
    \includegraphics[width=0.5\textwidth]{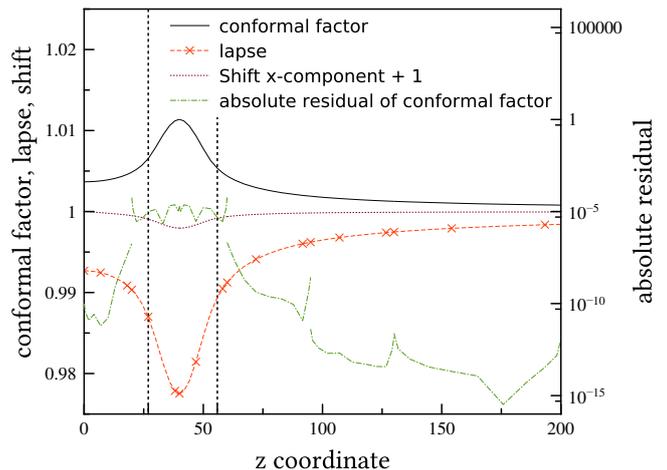}   
    \caption{
      \label{fig:BNSPsiAlphaBeta} 
      Steady state at a resolution of 11 collocation points per
      subpatch for the initial data of a binary neutron star system.
      We show data along the positive $z$-axis. The values on the
      negative axis are symmetric (anti-symmetric for the shift
      component $\beta^x$).  Solid line: conformal factor.  Orange
      dashed line with markers: lapse.  Purple dotted line:
      $x$-component of the shift, shown here with an offset of one for
      clarity.  Green dash-dotted line: absolute value of the residual
      for the conformal factor, as given by the right-hand side of
      Eq.~\eqref{eq:XCTSpsi}.  Vertical dashed line: position of the
      stellar surface.  }
\end{figure}
\section{Conclusions}\label{Section:Conclusions}

The most common types of relaxation methods are based on the famous
Gauss-Seidel method, which can be motivated by rewriting the problem
as a parabolic diffusion equation that relaxes from an initial guess
to the solution of the elliptic PDE.  In this paper we investigated a
new class of relaxation methods, which are not based on parabolic, but
rather on hyperbolic PDEs. In the literature hyperbolic relaxation is
usually discussed from the point of view that a hyperbolic equation is
given which may contain physical or numerical dissipation terms.  In
this work we assume that an elliptic equation is given, which is
extended to a hyperbolic relaxation equation for the purpose of
solving the elliptic equation. In some respect, hyperbolic relaxation
might actually be as well suited for the solution of elliptic PDEs as
parabolic relaxation.

We investigated how a hyperbolic relaxation method (HypRelax) can be
constructed for a general class of second order, non-linear elliptic
PDEs and discussed its structure and properties.  A discussion of the
general hyperbolicity properties has been carried out and three
specific choices for the relaxation were discussed. For the special
case of the Laplace equation, a mode analysis revealed that there is a
critical wavenumber at which the qualitative behavior of the modes is
changing. It has also been seen that at low wavenumber the damping
rate approaches that of the Jacobi method from above.  It is an
interesting topic for the future how the specific choice of the
relaxation system (choice of $b\indices{^j_i^\beta_\alpha}$) affects
the behavior of the modes, how that choice can be optimized, and what
would be the modes for more general elliptic equations.

Furthermore, we have shown how the standard types of boundary
conditions -- Dirichlet, Neumann and Robin boundary conditions -- can
be implemented within the hyperbolic relaxation framework.

With regard to advantages, hyperbolic relaxation shares with several
other methods the feature that it is ``matrix free'', \ie it is
possible to avoid the construction of an explicit matrix form by
applying the differential operator directly. Also note that as in
Jacobi methods, non-linear equations can be treated without
linearization, avoiding the additional work of \eg outer
Newton-Raphson iterations.

The ease of implementation is a key feature of hyperbolic relaxation.
In the past, we implemented a parallel geometric multigrid method for
the \texttt{BAM} and \texttt{Cactus}
codes~\cite{BraBru97,AlcBenBru00,BruGonHan06,MolMarJoh14}.  We also
interfaced \texttt{BAM} to the \texttt{hypre}
package~\cite{hypre_web,AnsBruTic04} for access to a variety of
elliptic solvers, in particular algebraic multigrid.  To avoid
complicated dependencies, some of the black hole initial data was
later computed with a stand-alone Newton-Raphson method, or even with
the direct solver in \texttt{MATLAB} for numerical
stability~\cite{DieBru13}. Our current production runs for a wide
range of binary neutron star initial data rely on the sophisticated
spectral solver \texttt{sgrid}~\cite{Tic09a,Tic12,DieMolJoh15}.
Unsurprisingly, compared to the various approaches just mentioned,
hyperbolic relaxation is very straightforward to implement given a
hyperbolic evolution code.

We have implemented the hyperbolic relaxation method in our spectral
hyperbolic evolution code \texttt{bamps}~\cite{Bru11,HilWeyBru15} (and
in a finite differencing test code).  For numerical tests we have
applied it to Poisson's equation, and we also presented applications
to numerical relativity initial data, where we have seen a high
robustness with respect to the choice of an initial guess and
regarding the simultaneous solution of the XCTS equations.  Given the
complexity of binary neutron star initial data and the correspondingly
involved numerical methods that have been developed to solve these
elliptic equations, \eg \cite{Tic17} and references therein, it is a
non-trivial result that the hyperbolic relaxation method results in a
robust and quite efficient elliptic solver.

We have seen that the damping rate at low wavenumbers is comparable to
that of the Jacobi method and thus very low.  In this study we were
most interested in the basic properties of the hyperbolic relaxation
method and thus we focussed on a simple scheme of successive
refinement, but there exist more sophisticated accelerators, in
particular multigrid methods, that could be implemented also for a
hyperbolic relaxation scheme. Another possible extension of our
hyperbolic relaxation implementation would be an adaptive mesh
refinement scheme, for which we see two main advantages.  First of
all, there are the usual savings due to optimized local resolution,
and secondly, since we want to use the solution of the elliptic
equation as initial data, this will provide us with a grid that should
already be well adapted to the evolution of the obtained data.
Another idea is to employ adaptive time stepping, for which however
special care would be required to keep the numeric scheme stable.

With regard to initial data for neutron stars, one of the potential
problems is the lack of differentiability at the surface of the stars.
We were prepared to evolve the neutron star data with a
high-resolution shock-capturing method, but in fact this was not
necessary given the strong diffusion of the equations. In applications
where the wave propagation feature is important, shock handling may be
a feature that comes at no extra cost assuming that there is an
evolution code providing the appropriate methods.

The investigations of this paper can only serve as a first step in the
exploration of this potentially promising branch of new relaxation
methods. Hyperbolic relaxation might become with further study, and in
particular in combination with acceleration methods, the basis of an
alternative numerical solution method for elliptic equations.

\begin{acknowledgments}

This work was supported in part by the DFG funded Graduiertenkolleg
1523/2, and the FCT (Portugal) IF Program IF/00577/2015. We thank
Niclas Moldenhauer and Tim Dietrich for fruitful discussions on the
construction of neutron star initial data.

\end{acknowledgments}
\appendix
\bibliography{hyp_relax.bbl}

\begin{thebibliography}{38}%
\makeatletter
\providecommand \@ifxundefined [1]{%
 \@ifx{#1\undefined}
}%
\providecommand \@ifnum [1]{%
 \ifnum #1\expandafter \@firstoftwo
 \else \expandafter \@secondoftwo
 \fi
}%
\providecommand \@ifx [1]{%
 \ifx #1\expandafter \@firstoftwo
 \else \expandafter \@secondoftwo
 \fi
}%
\providecommand \natexlab [1]{#1}%
\providecommand \enquote  [1]{``#1''}%
\providecommand \bibnamefont  [1]{#1}%
\providecommand \bibfnamefont [1]{#1}%
\providecommand \citenamefont [1]{#1}%
\providecommand \href@noop [0]{\@secondoftwo}%
\providecommand \href [0]{\begingroup \@sanitize@url \@href}%
\providecommand \@href[1]{\@@startlink{#1}\@@href}%
\providecommand \@@href[1]{\endgroup#1\@@endlink}%
\providecommand \@sanitize@url [0]{\catcode `\\12\catcode `\$12\catcode
  `\&12\catcode `\#12\catcode `\^12\catcode `\_12\catcode `\%12\relax}%
\providecommand \@@startlink[1]{}%
\providecommand \@@endlink[0]{}%
\providecommand \url  [0]{\begingroup\@sanitize@url \@url }%
\providecommand \@url [1]{\endgroup\@href {#1}{\urlprefix }}%
\providecommand \urlprefix  [0]{URL }%
\providecommand \Eprint [0]{\href }%
\providecommand \doibase [0]{http://dx.doi.org/}%
\providecommand \selectlanguage [0]{\@gobble}%
\providecommand \bibinfo  [0]{\@secondoftwo}%
\providecommand \bibfield  [0]{\@secondoftwo}%
\providecommand \translation [1]{[#1]}%
\providecommand \BibitemOpen [0]{}%
\providecommand \bibitemStop [0]{}%
\providecommand \bibitemNoStop [0]{.\EOS\space}%
\providecommand \EOS [0]{\spacefactor3000\relax}%
\providecommand \BibitemShut  [1]{\csname bibitem#1\endcsname}%
\let\auto@bib@innerbib\@empty
\bibitem [{\citenamefont {Saad}(2003)}]{Saa03}%
  \BibitemOpen
  \bibfield  {author} {\bibinfo {author} {\bibfnamefont {Y.}~\bibnamefont
  {Saad}},\ }\href@noop {} {\emph {\bibinfo {title} {Iterative Methods for
  Sparse Linear Systems}}},\ \bibinfo {edition} {2nd}\ ed.\ (\bibinfo
  {publisher} {SIAM Press},\ \bibinfo {address} {Philadelphia, USA},\ \bibinfo
  {year} {2003})\BibitemShut {NoStop}%
\bibitem [{\citenamefont {Saad}\ and\ \citenamefont {van~der
  Vorst}(2000)}]{SaaVor00}%
  \BibitemOpen
  \bibfield  {author} {\bibinfo {author} {\bibfnamefont {Y.}~\bibnamefont
  {Saad}}\ and\ \bibinfo {author} {\bibfnamefont {H.}~\bibnamefont {van~der
  Vorst}},\ }\href@noop {} {\bibfield  {journal} {\bibinfo  {journal} {J. Comp.
  App. Math.}\ }\textbf {\bibinfo {volume} {123}},\ \bibinfo {pages} {1}
  (\bibinfo {year} {2000})}\BibitemShut {NoStop}%
\bibitem [{\citenamefont {Hsiao}\ and\ \citenamefont {Liu}(1992)}]{HsiLiu92}%
  \BibitemOpen
  \bibfield  {author} {\bibinfo {author} {\bibfnamefont {L.}~\bibnamefont
  {Hsiao}}\ and\ \bibinfo {author} {\bibfnamefont {T.-P.}\ \bibnamefont
  {Liu}},\ }\href {\doibase 10.1007/BF02099268} {\bibfield  {journal} {\bibinfo
   {journal} {Communications in Mathematical Physics}\ }\textbf {\bibinfo
  {volume} {143}},\ \bibinfo {pages} {599} (\bibinfo {year}
  {1992})}\BibitemShut {NoStop}%
\bibitem [{\citenamefont {Nishihara}(1997)}]{Nis97}%
  \BibitemOpen
  \bibfield  {author} {\bibinfo {author} {\bibfnamefont {K.}~\bibnamefont
  {Nishihara}},\ }\href {\doibase http://dx.doi.org/10.1006/jdeq.1997.3268}
  {\bibfield  {journal} {\bibinfo  {journal} {Journal of Differential
  Equations}\ }\textbf {\bibinfo {volume} {137}},\ \bibinfo {pages} {384 }
  (\bibinfo {year} {1997})}\BibitemShut {NoStop}%
\bibitem [{\citenamefont {Wirth}(2014)}]{Wir14}%
  \BibitemOpen
  \bibfield  {author} {\bibinfo {author} {\bibfnamefont {J.}~\bibnamefont
  {Wirth}},\ }\href {\doibase 10.1016/j.jmaa.2014.01.034} {\bibfield  {journal}
  {\bibinfo  {journal} {Journal of Mathematical Analysis and Applications}\
  }\textbf {\bibinfo {volume} {414}},\ \bibinfo {pages} {666 } (\bibinfo {year}
  {2014})}\BibitemShut {NoStop}%
\bibitem [{\citenamefont {Alcubierre}\ \emph {et~al.}(2003)\citenamefont
  {Alcubierre}, \citenamefont {Br{\"u}gmann}, \citenamefont {Diener},
  \citenamefont {Koppitz}, \citenamefont {Pollney}, \citenamefont {Seidel},\
  and\ \citenamefont {Takahashi}}]{AlcBruDie02}%
  \BibitemOpen
  \bibfield  {author} {\bibinfo {author} {\bibfnamefont {M.}~\bibnamefont
  {Alcubierre}}, \bibinfo {author} {\bibfnamefont {B.}~\bibnamefont
  {Br{\"u}gmann}}, \bibinfo {author} {\bibfnamefont {P.}~\bibnamefont
  {Diener}}, \bibinfo {author} {\bibfnamefont {M.}~\bibnamefont {Koppitz}},
  \bibinfo {author} {\bibfnamefont {D.}~\bibnamefont {Pollney}}, \bibinfo
  {author} {\bibfnamefont {E.}~\bibnamefont {Seidel}}, \ and\ \bibinfo {author}
  {\bibfnamefont {R.}~\bibnamefont {Takahashi}},\ }\href@noop {} {\bibfield
  {journal} {\bibinfo  {journal} {Phys. Rev. D}\ }\textbf {\bibinfo {volume}
  {67}},\ \bibinfo {pages} {084023} (\bibinfo {year} {2003})},\ \Eprint
  {http://arxiv.org/abs/gr-qc/0206072} {gr-qc/0206072} \BibitemShut {NoStop}%
\bibitem [{\citenamefont {Balakrishna}\ \emph {et~al.}(1996)\citenamefont
  {Balakrishna}, \citenamefont {Daues}, \citenamefont {Seidel}, \citenamefont
  {Suen}, \citenamefont {Tobias},\ and\ \citenamefont {Wang}}]{BalDauSei96}%
  \BibitemOpen
  \bibfield  {author} {\bibinfo {author} {\bibfnamefont {J.}~\bibnamefont
  {Balakrishna}}, \bibinfo {author} {\bibfnamefont {G.}~\bibnamefont {Daues}},
  \bibinfo {author} {\bibfnamefont {E.}~\bibnamefont {Seidel}}, \bibinfo
  {author} {\bibfnamefont {W.-M.}\ \bibnamefont {Suen}}, \bibinfo {author}
  {\bibfnamefont {M.}~\bibnamefont {Tobias}}, \ and\ \bibinfo {author}
  {\bibfnamefont {E.}~\bibnamefont {Wang}},\ }\href@noop {} {\bibfield
  {journal} {\bibinfo  {journal} {Class. Quantum Grav.}\ }\textbf {\bibinfo
  {volume} {13}},\ \bibinfo {pages} {L135} (\bibinfo {year}
  {1996})}\BibitemShut {NoStop}%
\bibitem [{\citenamefont {Gustafsson}\ \emph {et~al.}(1995)\citenamefont
  {Gustafsson}, \citenamefont {Kreiss},\ and\ \citenamefont
  {Oliger}}]{GusKreOli95}%
  \BibitemOpen
  \bibfield  {author} {\bibinfo {author} {\bibfnamefont {B.}~\bibnamefont
  {Gustafsson}}, \bibinfo {author} {\bibfnamefont {H.-O.}\ \bibnamefont
  {Kreiss}}, \ and\ \bibinfo {author} {\bibfnamefont {J.}~\bibnamefont
  {Oliger}},\ }\href@noop {} {\emph {\bibinfo {title} {Time dependent problems
  and difference methods}}}\ (\bibinfo  {publisher} {Wiley},\ \bibinfo
  {address} {New York},\ \bibinfo {year} {1995})\BibitemShut {NoStop}%
\bibitem [{\citenamefont {Sarbach}\ and\ \citenamefont
  {Tiglio}(2012)}]{SarTig12}%
  \BibitemOpen
  \bibfield  {author} {\bibinfo {author} {\bibfnamefont {O.}~\bibnamefont
  {Sarbach}}\ and\ \bibinfo {author} {\bibfnamefont {M.}~\bibnamefont
  {Tiglio}},\ }\href {http://www.livingreviews.org/lrr-2012-9} {\bibfield
  {journal} {\bibinfo  {journal} {Living Reviews in Relativity}\ }\textbf
  {\bibinfo {volume} {15}} (\bibinfo {year} {2012})},\ \Eprint
  {http://arxiv.org/abs/1203.6443} {arXiv:1203.6443 [gr-qc]} \BibitemShut
  {NoStop}%
\bibitem [{\citenamefont {Hilditch}(2013)}]{Hil13}%
  \BibitemOpen
  \bibfield  {author} {\bibinfo {author} {\bibfnamefont {D.}~\bibnamefont
  {Hilditch}},\ }\href {\doibase 10.1142/S0217751X13400150} {\bibfield
  {journal} {\bibinfo  {journal} {Int. J. Mod. Phys.}\ }\textbf {\bibinfo
  {volume} {A28}},\ \bibinfo {pages} {1340015} (\bibinfo {year} {2013})},\
  \Eprint {http://arxiv.org/abs/1309.2012} {arXiv:1309.2012 [gr-qc]}
  \BibitemShut {NoStop}%
\bibitem [{\citenamefont {Br{\"u}gmann}(2013)}]{Bru11}%
  \BibitemOpen
  \bibfield  {author} {\bibinfo {author} {\bibfnamefont {B.}~\bibnamefont
  {Br{\"u}gmann}},\ }\href@noop {} {\bibfield  {journal} {\bibinfo  {journal}
  {J. Comput. Phys.}\ }\textbf {\bibinfo {volume} {235}},\ \bibinfo {pages}
  {216} (\bibinfo {year} {2013})},\ \Eprint {http://arxiv.org/abs/1104.3408}
  {arXiv:1104.3408 [physics.comp-ph]} \BibitemShut {NoStop}%
\bibitem [{\citenamefont {Hilditch}\ \emph {et~al.}(2016)\citenamefont
  {Hilditch}, \citenamefont {Weyhausen},\ and\ \citenamefont
  {Br{\"u}gmann}}]{HilWeyBru15}%
  \BibitemOpen
  \bibfield  {author} {\bibinfo {author} {\bibfnamefont {D.}~\bibnamefont
  {Hilditch}}, \bibinfo {author} {\bibfnamefont {A.}~\bibnamefont {Weyhausen}},
  \ and\ \bibinfo {author} {\bibfnamefont {B.}~\bibnamefont {Br{\"u}gmann}},\
  }\href {\doibase 10.1103/PhysRevD.93.063006} {\bibfield  {journal} {\bibinfo
  {journal} {Phys. Rev.}\ }\textbf {\bibinfo {volume} {D93}},\ \bibinfo {pages}
  {063006} (\bibinfo {year} {2016})},\ \Eprint
  {http://arxiv.org/abs/1504.04732} {arXiv:1504.04732 [gr-qc]} \BibitemShut
  {NoStop}%
\bibitem [{\citenamefont {Dain}(2006)}]{Dai04}%
  \BibitemOpen
  \bibfield  {author} {\bibinfo {author} {\bibfnamefont {S.}~\bibnamefont
  {Dain}},\ }\href@noop {} {\bibfield  {journal} {\bibinfo  {journal} {Lect.
  Notes Phys.}\ }\textbf {\bibinfo {volume} {692}},\ \bibinfo {pages} {117}
  (\bibinfo {year} {2006})},\ \Eprint {http://arxiv.org/abs/gr-qc/0411081}
  {gr-qc/0411081} \BibitemShut {NoStop}%
\bibitem [{\citenamefont {Kreiss}\ \emph {et~al.}(1998)\citenamefont {Kreiss},
  \citenamefont {Ortiz},\ and\ \citenamefont {Reula}}]{KreOrtReu98}%
  \BibitemOpen
  \bibfield  {author} {\bibinfo {author} {\bibfnamefont {H.-O.}\ \bibnamefont
  {Kreiss}}, \bibinfo {author} {\bibfnamefont {O.~E.}\ \bibnamefont {Ortiz}}, \
  and\ \bibinfo {author} {\bibfnamefont {O.~A.}\ \bibnamefont {Reula}},\ }\href
  {\doibase 10.1006/jdeq.1997.3341} {\bibfield  {journal} {\bibinfo  {journal}
  {Journal of Differential Equations}\ }\textbf {\bibinfo {volume} {142}},\
  \bibinfo {pages} {78 } (\bibinfo {year} {1998})}\BibitemShut {NoStop}%
\bibitem [{\citenamefont {Kreiss}\ and\ \citenamefont
  {Lorenz}(1998)}]{KreLor98}%
  \BibitemOpen
  \bibfield  {author} {\bibinfo {author} {\bibfnamefont {H.-O.}\ \bibnamefont
  {Kreiss}}\ and\ \bibinfo {author} {\bibfnamefont {J.}~\bibnamefont
  {Lorenz}},\ }\href {\doibase 10.1017/S096249290000283X} {\bibfield  {journal}
  {\bibinfo  {journal} {Acta Numerica}\ }\textbf {\bibinfo {volume} {7}},\
  \bibinfo {pages} {203} (\bibinfo {year} {1998})}\BibitemShut {NoStop}%
\bibitem [{\citenamefont {Hesthaven}(2000)}]{Hes00}%
  \BibitemOpen
  \bibfield  {author} {\bibinfo {author} {\bibfnamefont {J.~S.}\ \bibnamefont
  {Hesthaven}},\ }\href@noop {} {\bibfield  {journal} {\bibinfo  {journal}
  {Appl. Numer. Math.}\ }\textbf {\bibinfo {volume} {33}},\ \bibinfo {pages}
  {23} (\bibinfo {year} {2000})}\BibitemShut {NoStop}%
\bibitem [{\citenamefont {Hesthaven}\ \emph {et~al.}(2007)\citenamefont
  {Hesthaven}, \citenamefont {Gottlieb},\ and\ \citenamefont
  {Gottlieb}}]{HesGotGot07}%
  \BibitemOpen
  \bibfield  {author} {\bibinfo {author} {\bibfnamefont {J.~S.}\ \bibnamefont
  {Hesthaven}}, \bibinfo {author} {\bibfnamefont {S.}~\bibnamefont {Gottlieb}},
  \ and\ \bibinfo {author} {\bibfnamefont {D.}~\bibnamefont {Gottlieb}},\
  }\href@noop {} {\emph {\bibinfo {title} {Spectral Methods for Time-Dependent
  Problems}}}\ (\bibinfo  {publisher} {Cambridge University Press},\ \bibinfo
  {address} {Cambridge},\ \bibinfo {year} {2007})\BibitemShut {NoStop}%
\bibitem [{\citenamefont {Taylor}\ \emph {et~al.}(2010)\citenamefont {Taylor},
  \citenamefont {Kidder},\ and\ \citenamefont {Teukolsky}}]{TayKidTeu10}%
  \BibitemOpen
  \bibfield  {author} {\bibinfo {author} {\bibfnamefont {N.~W.}\ \bibnamefont
  {Taylor}}, \bibinfo {author} {\bibfnamefont {L.~E.}\ \bibnamefont {Kidder}},
  \ and\ \bibinfo {author} {\bibfnamefont {S.~A.}\ \bibnamefont {Teukolsky}},\
  }\href {\doibase 10.1103/PhysRevD.82.024037} {\bibfield  {journal} {\bibinfo
  {journal} {Phys.Rev.}\ }\textbf {\bibinfo {volume} {D82}},\ \bibinfo {pages}
  {024037} (\bibinfo {year} {2010})},\ \Eprint {http://arxiv.org/abs/1005.2922}
  {arXiv:1005.2922 [gr-qc]} \BibitemShut {NoStop}%
\bibitem [{\citenamefont {Gundlach}\ and\ \citenamefont
  {Mart{\'\i}n-Garc{\'\i}a}(2006)}]{GunGar05}%
  \BibitemOpen
  \bibfield  {author} {\bibinfo {author} {\bibfnamefont {C.}~\bibnamefont
  {Gundlach}}\ and\ \bibinfo {author} {\bibfnamefont {J.~M.}\ \bibnamefont
  {Mart{\'\i}n-Garc{\'\i}a}},\ }\href@noop {} {\bibfield  {journal} {\bibinfo
  {journal} {Class. Quantum Grav.}\ }\textbf {\bibinfo {volume} {23}},\
  \bibinfo {pages} {S387} (\bibinfo {year} {2006})},\ \Eprint
  {http://arxiv.org/abs/gr-qc/0506037} {gr-qc/0506037} \BibitemShut {NoStop}%
\bibitem [{\citenamefont {Ronchi}\ \emph {et~al.}(1996)\citenamefont {Ronchi},
  \citenamefont {Iacono},\ and\ \citenamefont {Paolucci}}]{RonIacPao96}%
  \BibitemOpen
  \bibfield  {author} {\bibinfo {author} {\bibfnamefont {C.}~\bibnamefont
  {Ronchi}}, \bibinfo {author} {\bibfnamefont {R.}~\bibnamefont {Iacono}}, \
  and\ \bibinfo {author} {\bibfnamefont {P.}~\bibnamefont {Paolucci}},\
  }\href@noop {} {\bibfield  {journal} {\bibinfo  {journal} {J. Comput. Phys.}\
  }\textbf {\bibinfo {volume} {124}},\ \bibinfo {pages} {93 } (\bibinfo {year}
  {1996})}\BibitemShut {NoStop}%
\bibitem [{\citenamefont {Boyd}(2001)}]{Boy01}%
  \BibitemOpen
  \bibfield  {author} {\bibinfo {author} {\bibfnamefont {J.~P.}\ \bibnamefont
  {Boyd}},\ }\href@noop {} {\emph {\bibinfo {title} {Chebyshev and Fourier
  Spectral Methods (Second Edition, Revised)}}}\ (\bibinfo  {publisher} {Dover
  Publications},\ \bibinfo {address} {New York},\ \bibinfo {year}
  {2001})\BibitemShut {NoStop}%
\bibitem [{\citenamefont {Press}\ \emph {et~al.}(2007)\citenamefont {Press},
  \citenamefont {Teukolsky}, \citenamefont {Vetterling},\ and\ \citenamefont
  {Flannery}}]{PreFlaTeu07a}%
  \BibitemOpen
  \bibfield  {author} {\bibinfo {author} {\bibfnamefont {W.~H.}\ \bibnamefont
  {Press}}, \bibinfo {author} {\bibfnamefont {S.~A.}\ \bibnamefont
  {Teukolsky}}, \bibinfo {author} {\bibfnamefont {W.~T.}\ \bibnamefont
  {Vetterling}}, \ and\ \bibinfo {author} {\bibfnamefont {B.~P.}\ \bibnamefont
  {Flannery}},\ }\href@noop {} {\emph {\bibinfo {title} {Numerical Recipes: The
  Art of Scientific Computing}}},\ \bibinfo {edition} {3rd}\ ed.\ (\bibinfo
  {publisher} {Cambridge University Press},\ \bibinfo {address} {New York,
  NY},\ \bibinfo {year} {2007})\BibitemShut {NoStop}%
\bibitem [{\citenamefont {Barrett}\ \emph {et~al.}(1993)\citenamefont
  {Barrett}, \citenamefont {Berry}, \citenamefont {Chan}, \citenamefont
  {Dongarra}, \citenamefont {Eijkhout}, \citenamefont {Romine},\ and\
  \citenamefont {van~der Vorst}}]{BarBerCha93}%
  \BibitemOpen
  \bibfield  {author} {\bibinfo {author} {\bibfnamefont {R.}~\bibnamefont
  {Barrett}}, \bibinfo {author} {\bibfnamefont {M.}~\bibnamefont {Berry}},
  \bibinfo {author} {\bibfnamefont {T.}~\bibnamefont {Chan}}, \bibinfo {author}
  {\bibfnamefont {J.}~\bibnamefont {Dongarra}}, \bibinfo {author}
  {\bibfnamefont {V.}~\bibnamefont {Eijkhout}}, \bibinfo {author}
  {\bibfnamefont {C.}~\bibnamefont {Romine}}, \ and\ \bibinfo {author}
  {\bibfnamefont {H.}~\bibnamefont {van~der Vorst}},\ }\href@noop {} {\emph
  {\bibinfo {title} {Templates for the Solution of Linear Systems: Building
  Blocks for Iterative Methods}}}\ (\bibinfo  {publisher} {SIAM},\ \bibinfo
  {address} {{\tt http://www.netlib.org/templates/}},\ \bibinfo {year}
  {1993})\BibitemShut {NoStop}%
\bibitem [{\citenamefont {Baumgarte}\ and\ \citenamefont
  {Shapiro}(2010)}]{BauSha10}%
  \BibitemOpen
  \bibfield  {author} {\bibinfo {author} {\bibfnamefont {T.~W.}\ \bibnamefont
  {Baumgarte}}\ and\ \bibinfo {author} {\bibfnamefont {S.~L.}\ \bibnamefont
  {Shapiro}},\ }\href@noop {} {\emph {\bibinfo {title} {Numerical Relativity:
  Solving {E}instein's Equations on the Computer}}}\ (\bibinfo  {publisher}
  {Cambridge University Press},\ \bibinfo {address} {Cambridge},\ \bibinfo
  {year} {2010})\BibitemShut {NoStop}%
\bibitem [{\citenamefont {Tichy}(2017)}]{Tic17}%
  \BibitemOpen
  \bibfield  {author} {\bibinfo {author} {\bibfnamefont {W.}~\bibnamefont
  {Tichy}},\ }\href {http://stacks.iop.org/0034-4885/80/i=2/a=026901}
  {\bibfield  {journal} {\bibinfo  {journal} {Reports on Progress in Physics}\
  }\textbf {\bibinfo {volume} {80}},\ \bibinfo {pages} {026901} (\bibinfo
  {year} {2017})}\BibitemShut {NoStop}%
\bibitem [{\citenamefont {Moldenhauer}(2012)}]{Mol12}%
  \BibitemOpen
  \bibfield  {author} {\bibinfo {author} {\bibfnamefont {N.}~\bibnamefont
  {Moldenhauer}},\ }\href@noop {} {\enquote {\bibinfo {title} {Initial data for
  neutron star binaries},}\ } (\bibinfo {year} {2012}),\ \bibinfo {note}
  {{M}aster {T}hesis, {U}niversity of {J}ena}\BibitemShut {NoStop}%
\bibitem [{\citenamefont {Moldenhauer}\ \emph {et~al.}(2014)\citenamefont
  {Moldenhauer}, \citenamefont {Markakis}, \citenamefont {Johnson-McDaniel},
  \citenamefont {Tichy},\ and\ \citenamefont {Br{\"u}gmann}}]{MolMarJoh14}%
  \BibitemOpen
  \bibfield  {author} {\bibinfo {author} {\bibfnamefont {N.}~\bibnamefont
  {Moldenhauer}}, \bibinfo {author} {\bibfnamefont {C.~M.}\ \bibnamefont
  {Markakis}}, \bibinfo {author} {\bibfnamefont {N.~K.}\ \bibnamefont
  {Johnson-McDaniel}}, \bibinfo {author} {\bibfnamefont {W.}~\bibnamefont
  {Tichy}}, \ and\ \bibinfo {author} {\bibfnamefont {B.}~\bibnamefont
  {Br{\"u}gmann}},\ }\href {\doibase 10.1103/PhysRevD.90.084043} {\bibfield
  {journal} {\bibinfo  {journal} {Phys. Rev. D}\ }\textbf {\bibinfo {volume}
  {90}},\ \bibinfo {pages} {084043} (\bibinfo {year} {2014})},\ \Eprint
  {http://arxiv.org/abs/1408.4136} {arXiv:1408.4136 [gr-qc]} \BibitemShut
  {NoStop}%
\bibitem [{\citenamefont {Tichy}(2009)}]{Tic09a}%
  \BibitemOpen
  \bibfield  {author} {\bibinfo {author} {\bibfnamefont {W.}~\bibnamefont
  {Tichy}},\ }\href {\doibase 10.1088/0264-9381/26/17/175018} {\bibfield
  {journal} {\bibinfo  {journal} {Classical Quantum Gravity}\ }\textbf
  {\bibinfo {volume} {26}},\ \bibinfo {pages} {175018} (\bibinfo {year}
  {2009})},\ \Eprint {http://arxiv.org/abs/0908.0620} {arXiv:0908.0620 [gr-qc]}
  \BibitemShut {NoStop}%
\bibitem [{\citenamefont {Bonazzola}\ \emph {et~al.}(1998)\citenamefont
  {Bonazzola}, \citenamefont {Gourgoulhon},\ and\ \citenamefont
  {Marck}}]{BonGouMar98a}%
  \BibitemOpen
  \bibfield  {author} {\bibinfo {author} {\bibfnamefont {S.}~\bibnamefont
  {Bonazzola}}, \bibinfo {author} {\bibfnamefont {E.}~\bibnamefont
  {Gourgoulhon}}, \ and\ \bibinfo {author} {\bibfnamefont {J.-A.}\ \bibnamefont
  {Marck}},\ }\href@noop {} {\bibfield  {journal} {\bibinfo  {journal} {Phys.
  Rev. D.}\ }\textbf {\bibinfo {volume} {58}},\ \bibinfo {pages} {104020}
  (\bibinfo {year} {1998})},\ \Eprint {http://arxiv.org/abs/astro-ph/9803086}
  {astro-ph/9803086} \BibitemShut {NoStop}%
\bibitem [{\citenamefont {Ansorg}(2007)}]{Ans06}%
  \BibitemOpen
  \bibfield  {author} {\bibinfo {author} {\bibfnamefont {M.}~\bibnamefont
  {Ansorg}},\ }\href {\doibase 10.1088/0264-9381/24/12/S01} {\bibfield
  {journal} {\bibinfo  {journal} {Classical Quantum Gravity}\ }\textbf
  {\bibinfo {volume} {24}},\ \bibinfo {pages} {S1} (\bibinfo {year} {2007})},\
  \Eprint {http://arxiv.org/abs/gr-qc/0612081} {arXiv:gr-qc/0612081 [gr-qc]}
  \BibitemShut {NoStop}%
\bibitem [{\citenamefont {Brandt}\ and\ \citenamefont
  {Br{\"u}gmann}(1997)}]{BraBru97}%
  \BibitemOpen
  \bibfield  {author} {\bibinfo {author} {\bibfnamefont {S.}~\bibnamefont
  {Brandt}}\ and\ \bibinfo {author} {\bibfnamefont {B.}~\bibnamefont
  {Br{\"u}gmann}},\ }\href@noop {} {\bibfield  {journal} {\bibinfo  {journal}
  {Phys. Rev. Lett.}\ }\textbf {\bibinfo {volume} {78}},\ \bibinfo {pages}
  {3606} (\bibinfo {year} {1997})},\ \Eprint
  {http://arxiv.org/abs/gr-qc/9703066} {gr-qc/9703066} \BibitemShut {NoStop}%
\bibitem [{\citenamefont {Alcubierre}\ \emph {et~al.}(2001)\citenamefont
  {Alcubierre}, \citenamefont {Benger}, \citenamefont {Br{\"u}gmann},
  \citenamefont {Lanfermann}, \citenamefont {Nerger}, \citenamefont {Seidel},\
  and\ \citenamefont {Takahashi}}]{AlcBenBru00}%
  \BibitemOpen
  \bibfield  {author} {\bibinfo {author} {\bibfnamefont {M.}~\bibnamefont
  {Alcubierre}}, \bibinfo {author} {\bibfnamefont {W.}~\bibnamefont {Benger}},
  \bibinfo {author} {\bibfnamefont {B.}~\bibnamefont {Br{\"u}gmann}}, \bibinfo
  {author} {\bibfnamefont {G.}~\bibnamefont {Lanfermann}}, \bibinfo {author}
  {\bibfnamefont {L.}~\bibnamefont {Nerger}}, \bibinfo {author} {\bibfnamefont
  {E.}~\bibnamefont {Seidel}}, \ and\ \bibinfo {author} {\bibfnamefont
  {R.}~\bibnamefont {Takahashi}},\ }\href@noop {} {\bibfield  {journal}
  {\bibinfo  {journal} {Phys. Rev. Lett.}\ }\textbf {\bibinfo {volume} {87}},\
  \bibinfo {pages} {271103} (\bibinfo {year} {2001})},\ \Eprint
  {http://arxiv.org/abs/gr-qc/0012079} {arXiv:gr-qc/0012079 [gr-qc]}
  \BibitemShut {NoStop}%
\bibitem [{\citenamefont {Br{\"u}gmann}\ \emph {et~al.}(2008)\citenamefont
  {Br{\"u}gmann}, \citenamefont {Gonz{\'a}lez}, \citenamefont {Hannam},
  \citenamefont {Husa}, \citenamefont {Sperhake},\ and\ \citenamefont
  {Tichy}}]{BruGonHan06}%
  \BibitemOpen
  \bibfield  {author} {\bibinfo {author} {\bibfnamefont {B.}~\bibnamefont
  {Br{\"u}gmann}}, \bibinfo {author} {\bibfnamefont {J.~A.}\ \bibnamefont
  {Gonz{\'a}lez}}, \bibinfo {author} {\bibfnamefont {M.}~\bibnamefont
  {Hannam}}, \bibinfo {author} {\bibfnamefont {S.}~\bibnamefont {Husa}},
  \bibinfo {author} {\bibfnamefont {U.}~\bibnamefont {Sperhake}}, \ and\
  \bibinfo {author} {\bibfnamefont {W.}~\bibnamefont {Tichy}},\ }\href
  {\doibase 10.1103/PhysRevD.77.024027} {\bibfield  {journal} {\bibinfo
  {journal} {Phys. Rev. D}\ }\textbf {\bibinfo {volume} {77}},\ \bibinfo
  {pages} {024027} (\bibinfo {year} {2008})},\ \Eprint
  {http://arxiv.org/abs/gr-qc/0610128} {arXiv:gr-qc/0610128 [gr-qc]}
  \BibitemShut {NoStop}%
\bibitem [{hyp()}]{hypre_web}%
  \BibitemOpen
  \href@noop {} {}\bibinfo {note} {{h}ypre -- High Performance Preconditioners:
  \\ {\tt http://www.llnl.gov/CASC/hypre/}}\BibitemShut {NoStop}%
\bibitem [{\citenamefont {Ansorg}\ \emph {et~al.}(2004)\citenamefont {Ansorg},
  \citenamefont {Br{\"u}gmann},\ and\ \citenamefont {Tichy}}]{AnsBruTic04}%
  \BibitemOpen
  \bibfield  {author} {\bibinfo {author} {\bibfnamefont {M.}~\bibnamefont
  {Ansorg}}, \bibinfo {author} {\bibfnamefont {B.}~\bibnamefont
  {Br{\"u}gmann}}, \ and\ \bibinfo {author} {\bibfnamefont {W.}~\bibnamefont
  {Tichy}},\ }\href@noop {} {\bibfield  {journal} {\bibinfo  {journal} {Phys.
  Rev.}\ }\textbf {\bibinfo {volume} {D70}},\ \bibinfo {pages} {064011}
  (\bibinfo {year} {2004})},\ \Eprint {http://arxiv.org/abs/gr-qc/0404056}
  {gr-qc/0404056} \BibitemShut {NoStop}%
\bibitem [{\citenamefont {Dietrich}\ and\ \citenamefont
  {Br{\"u}gmann}(2014)}]{DieBru13}%
  \BibitemOpen
  \bibfield  {author} {\bibinfo {author} {\bibfnamefont {T.}~\bibnamefont
  {Dietrich}}\ and\ \bibinfo {author} {\bibfnamefont {B.}~\bibnamefont
  {Br{\"u}gmann}},\ }\href@noop {} {\bibfield  {journal} {\bibinfo  {journal}
  {Phys.Rev.}\ }\textbf {\bibinfo {volume} {D89}},\ \bibinfo {pages} {024014}
  (\bibinfo {year} {2014})},\ \Eprint {http://arxiv.org/abs/1309.3087}
  {arXiv:1309.3087 [gr-qc]} \BibitemShut {NoStop}%
\bibitem [{\citenamefont {Tichy}(2012)}]{Tic12}%
  \BibitemOpen
  \bibfield  {author} {\bibinfo {author} {\bibfnamefont {W.}~\bibnamefont
  {Tichy}},\ }\href@noop {} {\bibfield  {journal} {\bibinfo  {journal} {Phys.
  Rev.}\ }\textbf {\bibinfo {volume} {D86}},\ \bibinfo {pages} {064024}
  (\bibinfo {year} {2012})},\ \Eprint {http://arxiv.org/abs/1209.5336}
  {arXiv:1209.5336 [gr-qc]} \BibitemShut {NoStop}%
\bibitem [{\citenamefont {Dietrich}\ \emph {et~al.}(2015)\citenamefont
  {Dietrich}, \citenamefont {Moldenhauer}, \citenamefont {Johnson-McDaniel},
  \citenamefont {Bernuzzi}, \citenamefont {Markakis}, \citenamefont
  {Br{\"u}gmann},\ and\ \citenamefont {Tichy}}]{DieMolJoh15}%
  \BibitemOpen
  \bibfield  {author} {\bibinfo {author} {\bibfnamefont {T.}~\bibnamefont
  {Dietrich}}, \bibinfo {author} {\bibfnamefont {N.}~\bibnamefont
  {Moldenhauer}}, \bibinfo {author} {\bibfnamefont {N.~K.}\ \bibnamefont
  {Johnson-McDaniel}}, \bibinfo {author} {\bibfnamefont {S.}~\bibnamefont
  {Bernuzzi}}, \bibinfo {author} {\bibfnamefont {C.~M.}\ \bibnamefont
  {Markakis}}, \bibinfo {author} {\bibfnamefont {B.}~\bibnamefont
  {Br{\"u}gmann}}, \ and\ \bibinfo {author} {\bibfnamefont {W.}~\bibnamefont
  {Tichy}},\ }\href@noop {} {\bibfield  {journal} {\bibinfo  {journal} {Phys.
  Rev.}\ }\textbf {\bibinfo {volume} {D92}},\ \bibinfo {pages} {124007}
  (\bibinfo {year} {2015})},\ \Eprint {http://arxiv.org/abs/1507.07100}
  {arXiv:1507.07100 [gr-qc]} \BibitemShut {NoStop}%
\end{thebibliography}%
\end{document}